\documentclass[12p]{iopart}
\usepackage{amssymb}  
\usepackage{subfig}
\usepackage{graphicx}
\usepackage{bm}
\usepackage{isotope}
\usepackage{cite}

\newcommand{\Nex}{N_\mathrm{ex}}
\newcommand{\Nmax}{N_\mathrm{max}}
\newcommand{\Nsex}{N_{\sigma,\mathrm{ex}}}

\renewcommand{\vec}[1]{\boldsymbol{\mathbf{\mathrm{#1}}}}
\newcommand{\uvec}[1]{\hat{\vec{#1}}}

\newcommand{\grpso}[1]{\mathrm{SO}(#1)}
\newcommand{\grpsu}[1]{\mathrm{SU}(#1)}
\newcommand{\grpu}[1]{\mathrm{U}(#1)}
\newcommand{\grpsptr}{\mathrm{Sp}(3,R)}

\begin{document}


\title{Rotational bands beyond the Elliott model}

\author{Ryan Zbikowski, 
Calvin W. Johnson}%
\address{%
Department of Physics, San Diego State University, San Diego, CA, 92182
}%
\address{%
Computational Science Research Center, San Diego State University, San Diego, CA, 92182
}%


\author{Anna E. McCoy}
\address{Institute for Nuclear Theory, University of Washington,
Seattle, Washington 98195-1550, USA}
\address{TRIUMF, Vancouver, British Columbia V6T 2A3, Canada}

\author{Mark A. Caprio, Patrick J. Fasano}
\address{Department of Physics, University of Notre Dame, Notre Dame,
Indiana 46556-5670, USA}

\date{\today}

\begin{abstract}
Rotational bands are commonplace in the spectra of atomic nuclei.  Inspired by early descriptions of these bands by 
quadrupole deformations of a liquid drop, 
Elliott constructed a discrete nucleon representations of $\mathrm{SU}(3)$ from fermionic creation and annihilation operators. Ever since,  Elliott's model has been foundational to 
descriptions of rotation in nuclei.  
Later work, however, suggested the symplectic extension $\mathrm{Sp}(3,R)$ 
provides a more unified picture.  We  decompose 
no-core shell-model nuclear wave functions into symmetry-defined subspaces for several beryllium isotopes, as well as $^{20}$Ne, using the quadratic 
Casimirs of both Elliott's $\mathrm{SU}(3)$ and $\mathrm{Sp}(3,R)$. 
The band structure,  delineated by strong $B(E2)$ values, has a more consistent description in $\mathrm{Sp}(3,R)$ rather than $\mathrm{SU}(3)$. 
{In particular, we confirm previous work finding in some nuclides strongly connected upper and lower 
bands with the same underlying symplectic structure.}
\end{abstract}

                              
                              \submitto{\jpg}

\maketitle

\section{Introduction}

Among the earliest pictures of the atomic nucleus was its description as a tiny drop of
dense fluid.  Not only did this metaphor motivate the semi-empirical mass
formula, it suggested interpreting nuclear excitation spectra as vibrations and
rotations of such a drop \cite{bohr1998nuclear2}.  While such a picture has many
limitations, not least the lack of discrete nucleons, its consequences
reverberate to this day.

To quantify the liquid drop model, the surface of the drop is parameterized in
terms of spherical coordinates $\theta$ and $\phi$ and expanded in spherical
harmonics: $R(\theta, \phi) = R_0 \sum_{l,m} a_{lm} Y_{lm} (\theta, \phi)$. In
the Bohr-Mottelson model, the coefficients $a_{lm}$ become dynamical
variables. The $l=0$ term is usually a constant (if dynamic, it relates to
breathing modes), and the $l=1$ terms are infinitesimal generators of
displacement. Hence the quadrupole $l=2$ terms are the first, and often the
last, important terms.  The five degrees of freedom from $m= -2, -1, 0, 1, 2$
correspond to the intrinsic shape deformation parameters $\beta, \gamma$ and
three Euler angles for orientation \cite{bohr1998nuclear2,ring2004nuclear,rowe2010nuclear}. The
dynamics of the quadrupole deformation can be described by five mass quadrupole
operators $\hat{Q}_{2 m} = r^2 Y_{2m}(\theta, \phi)$ plus the three components
of (orbital) angular momentum $\vec{L} = \vec{r} \times \vec{p}$
\cite{rosensteel1977shape}.  Together these eight operators form generators of
the rigid rotor group $\mathrm{ROT(3)}$~\cite{rowe1996dynamical}, or equivalently $[R^5] \mathrm{SO}(3)$~\cite{ui1970quantum,rosensteel1979algebraic,rowe2010fundamentals}.  This generator structure laid the foundation of describing nuclear
spectral bands using group theory \cite{talmi1993simple}.

The Bohr-Mottelson model represents the nucleus as an undifferentiated fluid.
Working under Giulio Racah, Phil Elliott~\cite{van2011scientific} considered an
alternative quadrupole operator $\hat{\mathcal{Q}}_{2m}$, one which is
equivalent to the mass quadrupole operator $\hat{Q}_{2 m}$ when restricted to a
single harmonic oscillator single shell, but which does not connect different
oscillator shells.  These $\hat{\mathcal{Q}}_{2m}$, taken together with the orbital angular momentum components
$\hat{L}_{1m}$, close under commutation.  Elliott
thereby arrived at a realization of the group $\mathrm{SU}(3)$~\cite{elliott1958:su3-part1,elliott1958:su3-part2,elliott1963:su3-part3,elliott1968:su3-part4,harvey1968nuclear}, generated by
operators which conserve the number of oscillator quanta, which made it appropriate 
for phenomenological calculations with a frozen core and valence particles restricted 
to a limited 
single particle space such as a single oscillator shell ~\cite{elliott1958:su3-part1,harvey1968nuclear}.
For detailed discussions
of the history and various models of rotation, see for example
\cite{rowe1996dynamical,carvalho1986symplectic}.

While $\mathrm{SU}(3)$ is not the only algebraic framework describing rotational
bands---indeed, it has been argued that the states of Elliott's $\mathrm{SU}(3)$
model are, under a technical definition, \textit{not} rotational
\cite{rosensteel1977shape} except in a certain limit
\cite{carvalho1986symplectic}---it is one of the best known and most widely
referenced.  A central question is whether an $\mathrm{SU}(3)$ picture is
sufficient. Because Elliott's $\mathrm{SU}(3)$ model, by design, is restricted
to states lying in the valence space, one can ask whether or not rotational
bands extend to states outside the valence space
\cite{goode1970ground,ap-126-1980-343-Rosensteel,draayer1984symplectic,PhysRevC.91.014310}.
A long-favored and natural alternative extension to $\mathrm{SU}(3)$ is the
symplectic extension $\mathrm{Sp}(3,R)$
\cite{PhysRevLett.38.10,rowe1996dynamical,PhysRevLett.98.162503,PhysRevC.76.014315,0954-3899-35-12-123101},
which includes the kinetic energy and the mass quadrupole operator $\hat{Q}_{2 m}$
(which, unlike the quadrupole operator of
Elliott's $\mathrm{SU}(3)$ model, is not restricted to a single shell).
 
In either an $\mathrm{SU}(3)$ or an $\mathrm{Sp}(3,R)$ rotational model, the
rotational bands lie within irreducible representations (irreps), which are
subspaces invariant under the generators of the respective group.  Such a
description of nuclear rotations is at best approximate, as realistic nuclear
interactions break the invariance, mixing wavefunctions across irreps.
Nonetheless, $\mathrm{SU}(3)$ and $\mathrm{Sp}(3,R)$ have been observed to be
good approximate symmetries of
nuclei~\cite{PhysRevLett.98.162503,0954-3899-35-12-123101,0954-3899-35-9-095101,ppnp-67-2012-516-Draayer,PhysRevLett.111.252501,dytrych2016efficacy,launey2016symmetry,prl-124-2020-042501-Dytrych,PhysRevC.76.014315,draayer2011ab,mccoy2018symplectic,mccoy2018:diss,mccoy2020emergent}
with the majority of the wavefunction spread over a relatively small number of
irreps. 
{Indeed, this is the motivation of calculations 
 carried out in bases made up of irreps of a chosen 
{algebra}, for example 
 SU(3)~\cite{PhysRevLett.111.252501,dytrych2016efficacy} 
or $\mathrm{Sp}(3R)$~\cite{mccoy2018symplectic,mccoy2020emergent},} which can  be truncated to include only the most ``important"
irreps.

In this paper we carry out large-basis no-core shell model (NCSM) calculations of selected
light nuclei {using a realistic, microscopic interaction~\cite{machleidt2011chiral}}, and decompose their underlying algebraic structure, directly
comparing $\mathrm{SU}(3)$ and $\mathrm{Sp}(3,R)$. It is important to emphasize
that the NCSM calculations are done without any assumptions of underlying group
structure. To examine the anatomy of rotational band members, which are
recognized by strong $E2$ transitions, we partition the wave functions into
subspaces spanned by irreps with the same quantum numbers and find the fraction
of each wave function contributed by each of these subspaces.  We accomplish
this by noting that these subspaces are eigenspaces of the Casimir operator of
the relevant group, which allows us to use an efficient decomposition
technique~\cite{PhysRevC.63.014318,PhysRevC.91.034313,PhysRevC.95.024303}.
Rotational bands in NCSM calculations for the beryllium isotopes have been
discussed in detail in
references~\cite{caprio2013emergence,PhysRevC.91.014310,caprio2015collective,caprio2019ab,caprio2020probing}.
In this work, we focus on a selection of these $p$-shell nuclides
($\isotope[7-10]{Be}$), as well as the $sd$-shell nucleus $\isotope[20]{Ne}$.

We find that while both $\mathrm{SU}(3)$ and
$\mathrm{Sp}(3,R)$ do provide an approximate descriptions of the rotational
bands, $\mathrm{Sp}(3,R)$ provide a more consistent description than
$\mathrm{SU}(3)$, in particular by including band members missed by $\mathrm{SU}(3)$. Furthermore, our work supports and extends a recent finding
\cite{mccoy2020emergent} that in some nuclei there exists an upper, excited
band-like structure with strong $E2$ transitions to the ground-state band,
both bands sharing a common $\mathrm{Sp}(3,R)$ description.  In fact, rather 
than being viewed as two separate but related bands, on the basis of our results 
we believe both ``bands'' should be considered part of a unified whole.

\section{Models and methods}

Physicists often refer to rotational and vibrational motion, but these are
pictures taken from classical physics.  Identifying analogous behavior in
quantum systems, especially in complex many-body systems, is not trivial.  One
way to define a rotational band is as a set of states with distinct angular
momentum quantum numbers, all projected from the same intrinsic
state~\cite{bohr1998nuclear2,ring2004nuclear,rowe2010nuclear}.  The wave function then
factorizes into the product of a rotational wave function and an intrinsic wave
function describing the structure in the body-fixed frame.  (Such a body-fixed
frame is only defined if the nucleus has quadrupole
moments with small variances~\cite{rosensteel1977shape}.)  This assumption
leads to specific predictions for ratios of electric quadrupole moments and $E2$
transition strengths.

The most commonly observed case is of an axially symmetric instrinsic state.
Here each band member has definite angular momentum $J$ \emph{and} definite
projection $K$ of $J$ onto the intrinsic symmetry axis.  The projection $K$ is
determined by the intrinsic state and is thus common to all band members, yielding
a band consisting of states with angular momentum $J\ge K$.  The energies of the
band members are given by
\begin{equation}
  \label{eqn:E-rotor}
    E(J)=E_K+A[J(J+1)-K^2+\delta_{K,\frac12}a(-1)^{J+1/2}(J+1/2)],
\end{equation}
where the rotational energy constant $A\equiv \hbar^2/(2\mathcal{I})$ is
inversely proportional to the moment of inertia $\mathcal{I}$, and $E_K$ is the
energy of the intrinsic state.  The final term in brackets, which contributes
only for $K=1/2$, arises from the Coriolis interaction and introduces an energy
staggering between alternate band members.

Rotational bands are empirically confirmed  through enhanced electric quadrupole
transition strengths, or $B(E2)$s, among band members, where the overall scale
of these is determined by the quadrupole moment of the intrinsic
state~\cite{bohr1998nuclear2,rowe2010nuclear,ring2004nuclear}.
While $E2$ transitions may readily identify 
 band members, the mere presence of
rotational $E2$ patterns provides no insight into the underlying intrinsic
structure.  Elliott's algebraic rotational model provided a first link between
the collective rotational model and the microscopic shell
model~\cite{elliott1958:su3-part1,elliott1958:su3-part2,elliott1963:su3-part3,elliott1968:su3-part4,harvey1968nuclear}, but it is limited by the
inherent assumption that the intrinsic state has definite $\mathrm{SU}(3)$
symmetry.  In this work, we carry out our investigation in a shell model (or,
more specifically, NCSM) framework, which makes no assumption of underlying
symmetries.  We review this shell model framework in section~\ref{sm}.
Section~\ref{dynamical} provides a brief introduction to the groups and their
associated algebras which underpin the dynamical symmetries we are looking for
in the nuclear system.  To investigate the symmetries of members of our
calculated rotational bands, we decompose them by partitioning the wave
functions into subspaces defined by the symmetry, as described in section~\ref{whichsymmetry}.

\subsection{The shell model context}

\label{sm}

We carry out our investigation in the context of the configuration-interaction
method: the nuclear wave function is expanded in a basis,
\begin{equation}
| \Psi \rangle = \sum_\alpha v_\alpha | \alpha \rangle,
\end{equation}
where the basis states are, for example, antisymmetrized products of
single-particle states or Slater determinants, or their representation in
occupation space, $ | \alpha \rangle = \hat{a}^\dagger_1 \hat{a}^\dagger_2
\hat{a}^\dagger_3 \ldots \hat{a}^\dagger_A | 0 \rangle$. Such basis states have
a discrete, finite number of fermions, and with a sufficiently large basis one
can in principle describe any nucleus \cite{BG77,br88,ca05}.

In configuration-interaction calculations one uses single particle states with
good angular momentum, that is, eigenstates of the total squared angular
momentum operator $\hat{J}^2$ and $z$-component $\hat{J}_z$.  Often one uses a
harmonic oscillator single particle basis, in part because matrix elements of
translationally invariant operators, including the Hamiltonian, are then
straightforward to compute, and in part because this also allows one to
rigorously address center-of-mass motion~\cite{caprio2020:intrinsic}, concerns
peripheral to our work here.  The single-particle harmonic oscillator
Hamiltonian $\hat{H}_0 = - \frac{\hbar^2}{2m} \nabla^2 + \frac{1}{2} m \omega^2
r^2$ has eigenvalues $\hbar \omega ( N + 3/2)$, where $N\equiv 2n+l$ is the
principal quantum number, in terms of the orbital angular momentum $l$ and the
number of nodes $n$ in the radial wave function. In common usage the distinction
between \textit{orbitals} and \textit{shells} can be vague.  We define
\textit{orbitals} as states with a specified $n, l,$ and $j$, while
\textit{shells} (or \textit{major oscillator shells}) are defined by $N$.

In empirical many-body spaces, such as those addressed originally by the Elliott
model, one has a frozen, inert core, with all orbitals occupied up to some
maximum, and the valence or active space is a fixed set of orbitals, often a
major oscillator shell, such as the $p$-, $sd$- or $pf$-shells.

In contrast, \textit{no-core} or \textit{ab initio} shell model (NCSM)
calculations allow particles to be excited in and out of many shells
\cite{navratil2000large,barrett2013ab}.  The model space is usually 
specified as follows: for any Slater determinant (or its occupation-space
representation) with fixed numbers of protons and neutrons, let $N_A = \sum_i
N_i$, where $N_i$ is the principal quantum number for the $i$th particle. That
is, $N_A$ is the total number of oscillator quanta in the many-body state, and,
when applied to an $A$-body state, the eigenvalue of $\hat{H}_0$ is $\hbar
\omega (N_A + 3A/2)$.  There will be some minimum value of $N_A$, $N_0$,
dictated by the Pauli principle.  One often labels
configurations not by their absolute $N_A$ but rather by the number of
excitation quanta $\Nex=N_A-N_0$ above the minimum.  These configurations are
also called $\Nex\hbar \omega$ excitations (i.e., $0\hbar\omega$, $2\hbar
\omega$, etc.), in reference to the corresponding oscillator energy.

The truncation of the no-core shell model many-body space is specified by
$N_\mathrm{max}$, which is the maximal $\Nex$ for the basis configurations (that
is, the difference between the minimal and maximal allowed values of $N_A$).  To
fully define the space, one must also specify the frequency of the
single-particle harmonic oscillator basis, conventionally quoted as the
oscillator energy $\hbar \omega$.  Because the parity for a single-particle
orbital is $\pi = (-1)^l=(-1)^N$, configurations with even $\Nex$
($\Nex=0,2,4,\ldots$) have ``natural'' parity (\textit{i.e.}, that of the lowest
Pauli allowed filling), while configurations with odd $\Nex$
($\Nex=1,3,5,\ldots$) have the opposite, ``unnatural'' parity.

The $\Nmax=0$ space, consisting of configurations with $\Nex = 0$, corresponds
to the typical valence model space of  empirical shell model calculations carried out in
a single oscillator shell. Here, in order to connect our multi-shell calculations 
with such empirical calculations, we will often refer to the $\Nex=0$ subspace as the 
valence subspace.
For example, in the beryllium isotopes we consider
below, the valence space consists of states with filled $0s_{1/2}$ orbital (that
is, a $^4$He core) with valence protons and neutrons restricted to the $0p$
shell, while for $^{20}$Ne the valence space has a $^{16}$O core with valence
nucleons in the $1s$-$0d$ shell.  Most low-lying states in NCSM calculations
have a dominant fraction of the wave function in the valence subspace.
 
 \begin{figure}[h]
 \centering
	\includegraphics[width=0.9\linewidth]{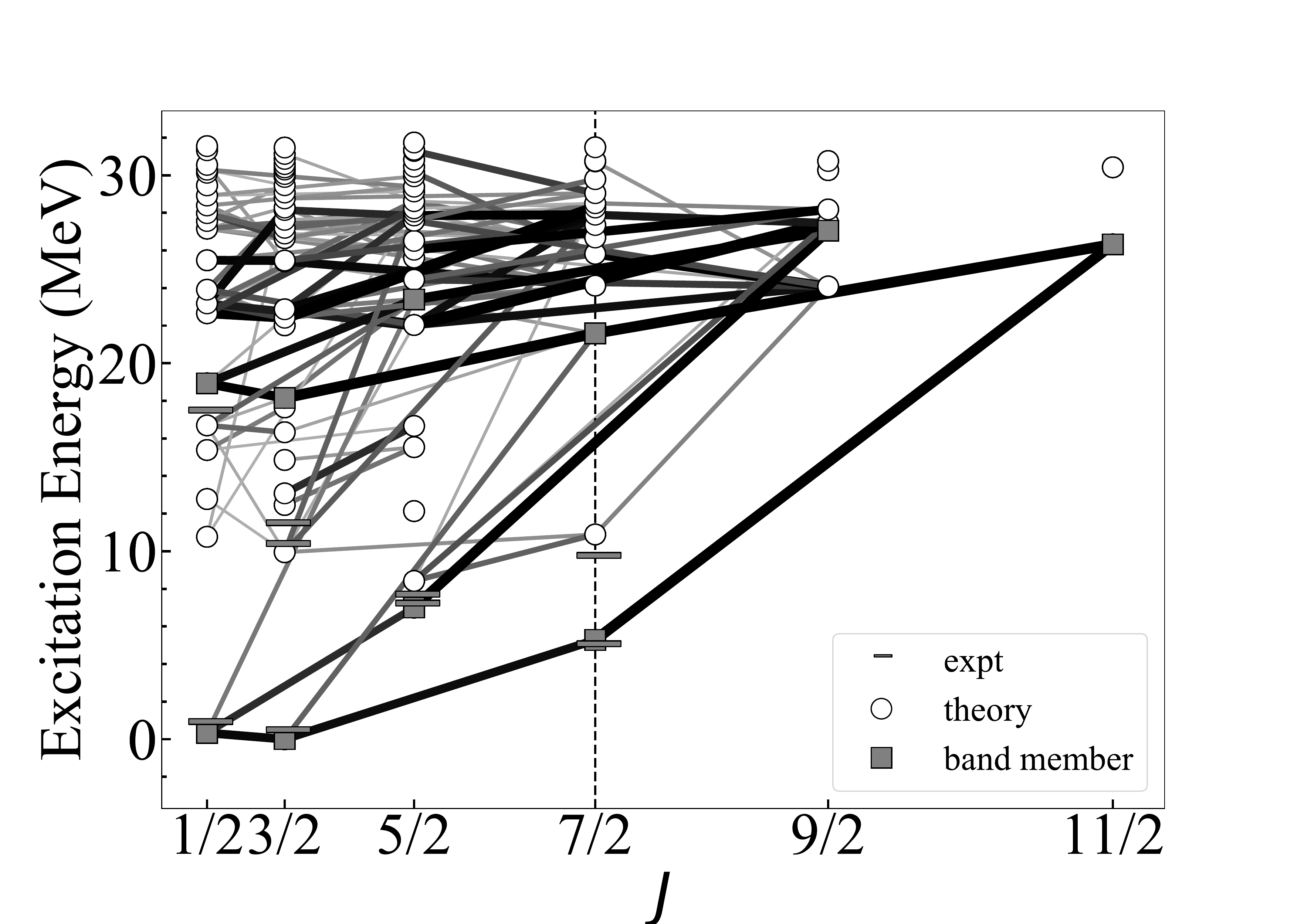}
	\caption{Excitation spectrum of the natural, negative parity states of
          ${}^{7}$Be, computed at $N_\mathrm{max} =10$ with a single-particle
          harmonic oscillator basis frequency $\hbar \omega = 20$ MeV, using a
          two-body chiral force at N3LO. Experimental energies are presented as
          shaded bars are shown for comparison (see text). Rotational band
          members are indicated using shaded squares.  These states are decomposed into
          group irreps in Figs.~\ref{fig:be7decomp_a} and~\ref{fig:be7decomp_b}. `Downward' (specifically
          from higher $J$ to lower $J$) $E2$ transitions connecting to band members
          are indicated with solid lines. Line thickness and shading
          is proportional to $B(E2)$ strength. Only strengths above 1.5
          $e^2\mathrm{fm}^4$ are shown. The vertical dashed line denotes the
          maximal angular momentum within the lowest harmonic oscillator
          configuration, that is, the valence subspace.  }
	\label{fig:be7be2}
\end{figure}
 
NCSM calculations use realistic nucleon-nucleon interactions, by which we mean
interactions fitted primarily to two- and three-body data such as scattering
phase shifts and deuteron properties, which are then applied to $A$-body systems
such as we describe herein.  Here we used a chiral
next-to-next-to-next-to-leading order (N3LO) interaction
\cite{machleidt2011chiral}.  We evolved the interaction via the similarity renormalization
group \cite{PhysRevC.75.061001} to a standard value of $\lambda= 2.0$
fm$^{-1}$. We did not use three-body forces, although their effect would be
interesting in future work.

For the many-body calculation we used the {\tt
  BIGSTICK} code \cite{BIGSTICK,johnson2018bigstick}, which solves the large,
sparse, many-body eigenvalue problem via the Lanczos algorithm \cite{Whitehead},
and which uses an $M$-scheme basis, that is, the many-body basis fixes the total
$M$ (or eigenvalue of or $\hat{J}_z$), and the overall parity, but no other
quantum numbers.  We carried out calculations of $^7$Be and $^8$Be at $N_\mathrm{max}=10$,
  while the `unnatural' parity states of $^9$Be we computed at
  $N_\mathrm{max}=9$, and $^{10}$Be
  we computed in a $N_\mathrm{max}=8$ space. Finally, we studied $^{20}$Ne in an
  $N_\mathrm{max}=4$ space.  These calculations had
  $M$-scheme basis dimensions between a few tens of millions and a few hundreds
  of millions.  All of our calculations used a harmonic oscillator
  basis frequency of $\hbar \omega= 20$ MeV.

Our guiding example in this paper is $^7$Be. Fig.~\ref{fig:be7be2} plots the
excitation energies of low-lying negative (natural) parity states of $^7$Be,
versus angular momentum $J$, with $J$ spaced as $J(J+1)$, as appropriate for
rotational analysis.  We also plot the downward ($J$-decreasing) $B(E2)$
strengths, computed using the standard operator \cite{BG77} with bare charges,
as solid lines, with the width and shading approximately proportional to the log
strength. The weakest transitions are omitted, for visual clarity.  NCSM $B(E2)$
values are known to be sensitive both to the choice of $N_\mathrm{max}$ and of
the oscillator parameter $\hbar \omega$ for the single-particle wave functions.
Nonetheless, the relative strengths of the $B(E2)$ values are comparatively
insensitive to these choices and are therefore useful in identifying members of
a band.  (See, \textit{e.g.}, references~\cite{caprio2019ab,caprio2020probing}
for convergence studies of NCSM calculations for the nuclei considered here.)
For example, here for $^7$Be, from the relative $E2$ strengths in
Fig.~\ref{fig:be7be2} it can be seen that, while most of the yrast states belong
to a clearly identified $K=1/2$ ground-state band, the yrast $9/2^-_1$ is not part of
the ground-state band; rather it is the $9/2^-_2$ state that is strongly
connected to the band.

\subsection{Dynamical symmetries and group theoretical framework}

\label{dynamical}

Symmetries and, in particular, dynamical symmetries provide valuable insight
into the nature of nuclear states and the nuclear excitation
spectrum~\cite{hecht1973:nuclear-symmetries,wybourne1974:groups,talmi1993simple,chen1989group,iachello1994:dynsymm,rowe2010fundamentals,iachello2015:liealg}.
Dynamical symmetries arise by construction in algebraic
models~\cite{rowe2010fundamentals}, which provide a simplified description of
nuclear dynamics, yet they can persist in the full nuclear many-body problem
with realistic nuclear forces.

Most familiar, perhaps, is the case in which the Hamiltonian is invariant under
a set of transformations that define the symmetry
group~\cite{rowe2010fundamentals} for the problem.  For example, the
rotationally invariant nuclear many-body problem has as its symmetry group
$\grpso{3}$ or, since fermions are involved, $\grpsu{2}$.  If the Hamiltonian is
invariant with respect to symmetry transformations, then the symmetry group can
only transform an eigenstate into other degnerate eigenstates, thereby forming a
degnerate multiplet.  Thus it is convenient to classify eigenstates into
irreducible representations (\textit{irreps}) of the symmetry group.  Irreps are
subspaces (of the full Hilbert space) which are invariant under the group
transformations and which cannot be reduced further, \textit{i.e.}, broken down
into smaller invariant subspaces.

The irreps of a symmetry group are labeled by a 
quantum number or numbers $\Gamma$.  
 The states within an irrep can be labeled by some
additional quantum number(s) $i$, as $|\Gamma
i\rangle$~\cite{wybourne1974:groups}.  For the familiar case of $\grpsu{2}$, the
irreps are the angular momentum multiplets $| J M \rangle$.

Often many irreps in a Hilbert space share the same
$\Gamma$, requiring additional quantum numbers denoted generically by
$\alpha$~\cite{wybourne1974:groups,talmi1993simple,rowe2010fundamentals}, so
that the states are $|\alpha\Gamma i\rangle$.  For instance, the space for a
rotationally invariant problem will contain many states of the same angular
momentum $J$, and the eigenstates are fully labeled as $| \alpha J M \rangle$.

While states naturally form irreps when the Hamiltonian is invariant under a
symmetry group, they may also be organized into irreps in the more general case
of dynamical symmetry.  A simple, classic example of dynamical symmetry, relevant to the present investigation, can be 
constructed in Elliott's $\grpsu{3}$
model~\cite{elliott1958:su3-part1,elliott1958:su3-part2,elliott1963:su3-part3,elliott1968:su3-part4,harvey1968nuclear} with the right 
 choice of interaction within the nuclear shell model.  

Elliot's $\grpsu{3}$ group was constructed with deformation and rotation in mind, 
and with generators chosen so that the group transformations do not excite nucleons  between
major shells.  
The generators include special quadrupole operators
\begin{equation} \label{ElliottQ}
    \hat{\mathcal{Q}}_{2m} = \sum_{i=1}^A \sqrt{\frac{4\pi}{5}} \left (\frac{r_i^{2}}{b^{2}}Y_{2m}
    (\uvec{r}_i) + b^{2}p_i^{2}Y_{2m}(\uvec{p}_i) \right ),
\end{equation}
where $b$ is the shell model oscillator length parameter.  Unlike the usual (mass)
quadrupole operators $\hat{Q}_{2m}$, the Elliott
quadrupole operators $\hat{\mathcal{Q}}_{2m}$ conserve the number of oscillator quanta;
yet within a single shell they are simply proportional to the usual
$\hat{Q}_{2m}$.   The remaining generators are the orbital angular
momentum operators $\hat{L}_{1m}=\sum_i \hat{L}_{1m,i}$, which by themselves
generate the group $\grpso{3}$ of rotations on the coordinate degrees of
freedom.  Elliott chose the definition~(\ref{ElliottQ}) so as to ensure that the
$\hat{\mathcal{Q}}_{2}$ and $\hat{L}_1$ operators close under commutation, and
have the commutator structure of $\grpsu{3}$ group generators.  We  thus have 
a subgroup chain $\grpsu{3}\supset\grpso{3}$.

In Elliott's $\grpsu{3}$ framework a natural shell model Hamiltonian
 is~\cite{harvey1968nuclear}
\begin{equation}
  \label{eqn:HQQ}
H=-\chi
\mathcal{Q}\cdot\mathcal{Q},
\end{equation}
where the dot represents the standard spherical tensor scalar
product~\cite{talmi1993simple}.  In this model, rotational bands emerge from
$\grpsu{3}\supset\grpso{3}$ irreps. An irrep of $\grpsu{3}$ is labeled by two
quantum numbers $(\lambda, \mu)$.  Depending on the  values of $(\lambda, \mu)$, the $\grpsu{3}$
irrep will contain several $\grpso{3}$ irreps, that is, states having different
values $L$ of the total orbital angular momentum.  These states naturally
organize into one or more rotational bands~\cite{harvey1968nuclear}, 
\textit{e.g.}, a $(4,2)$ irrep contains states with $L=0,2,4$
($K=0$) and $L=2,3,4,5,6$ ($K=2$).  The total orbital angular momentum $L$ and
the total spin $S$ combine to a  total angular momentum $J$.  The
simple Elliott Hamiltonian~(\ref{eqn:HQQ}) gives
bands with energies depending only on the orbital angular momentum, as $L(L+1)$, 
but  Elliott and Wilsdon~\cite{elliott1968:su3-part4} showed the
microscopic spin-orbit interaction mixes states of
different $L$ (but the same $J$) to give rotational bands with the familiar
rotational dependence~(\ref{eqn:E-rotor}) of energies on $J$.

A natural extension of Elliott's $\grpsu{3}$ group is the symplectic group
$\grpsptr$~\cite{PhysRevLett.38.10,rowe1996dynamical,PhysRevLett.98.162503,PhysRevC.76.014315,0954-3899-35-12-123101}
in three dimensions.  In addition to the $\grpsu{3}$ generators
$\hat{\mathcal{Q}}_{2m}$ and $\hat{L}_{1m}$, the generators of this group
include the harmonic oscillator Hamiltonian $\hat{H}_0$ and symplectic raising
and lowering (or ladder) operators $\hat{A}_{lm}$ and
$\hat{B}_{lm}$, with $l=0$ and $2$.  These ladder operators physically represent creation
and annihilation operators, respectively, for giant monopole and quadrupole
resonances.  Unlike the $\grpsu{3}$ generators, which conserve the number of
oscillator quanta, the symplectic raising and lowering operators add or
remove two quanta, respectively, to the nuclear many-body state.  They thus
connect $\Nex \hbar\omega$ shell model spaces differing by $2$ in $\Nex$.

If $\grpsptr$ dynamical symmetry holds for the nucleus, states will be organized
into $\grpsptr$ irreps.  Here we have the subgroup chain $\grpsptr\supset
\grpu{3} \sim \grpu{1}\times \grpsu{3}\supset\grpso{3}$.  The $\grpu{3}$ group
here is simply obtained by combining Elliott's $\grpsu{3}$ with the trivial
$\grpu{1}$ group of the harmonic oscillator Hamiltonian $\hat{H}_0$.  The set of states
in a $\grpu{3}$ irrep (\textit{i.e.}, the invariant subspace) is the same as for an Elliott $\grpsu{3}$ irrep, but the
new operator $\hat{H}_0$ provides a further label~--- the number of oscillator quanta or,
equivalently, $\Nex$~--- beyond the usual $\grpsu{3}$ labels
$(\lambda,\mu)$.
The number of
oscillator quanta is relevant in the context of $\grpsptr$ dynamical symmetry, as an $\grpsptr$ irrep spans many, in
fact, infinitely many, $\Nex \hbar \omega$ spaces. One builds an $\grpsptr$
irrep $\sigma=\Nsex(\lambda_\sigma,\mu_\sigma)$  recursively,  starting
from one Elliot $\grpsu{3}$ irrep $(\lambda_\sigma,\mu_\sigma)$, at some lowest
number $\Nsex$ of oscillator excitations.  Then by repeatedly laddering with the $\grpsptr$ raising operator $\hat{A}$ one obtains an infinite tower of $\grpu{3}$ irreps 
with different $\Nex=\Nsex,\Nsex+2,\Nsex+4$, and so on.

When one moves beyond simple algebraic models, such as Elliott's for $\grpsu{3}$ or analogous
algebraic models for $\grpsptr$~\cite{rowe2016:micsmacs}, the interaction may 
break the dynamical symmetry and
mix the wave function across irreps.  While one might expect the
mixing to be state dependent, surprisingly often one finds similar patterns of
mixing across multiple states~\cite{rochford1988survival,PhysRevC.63.014318}.
This is \textit{quasi-dynamical} symmetry~\cite{PhysRevC.58.1539,rowe2000quasi,bahri20003}.

\subsection{But which symmetry?}

\label{whichsymmetry}

\begin{figure}[h]
\centering
	\includegraphics[width=0.9\linewidth]{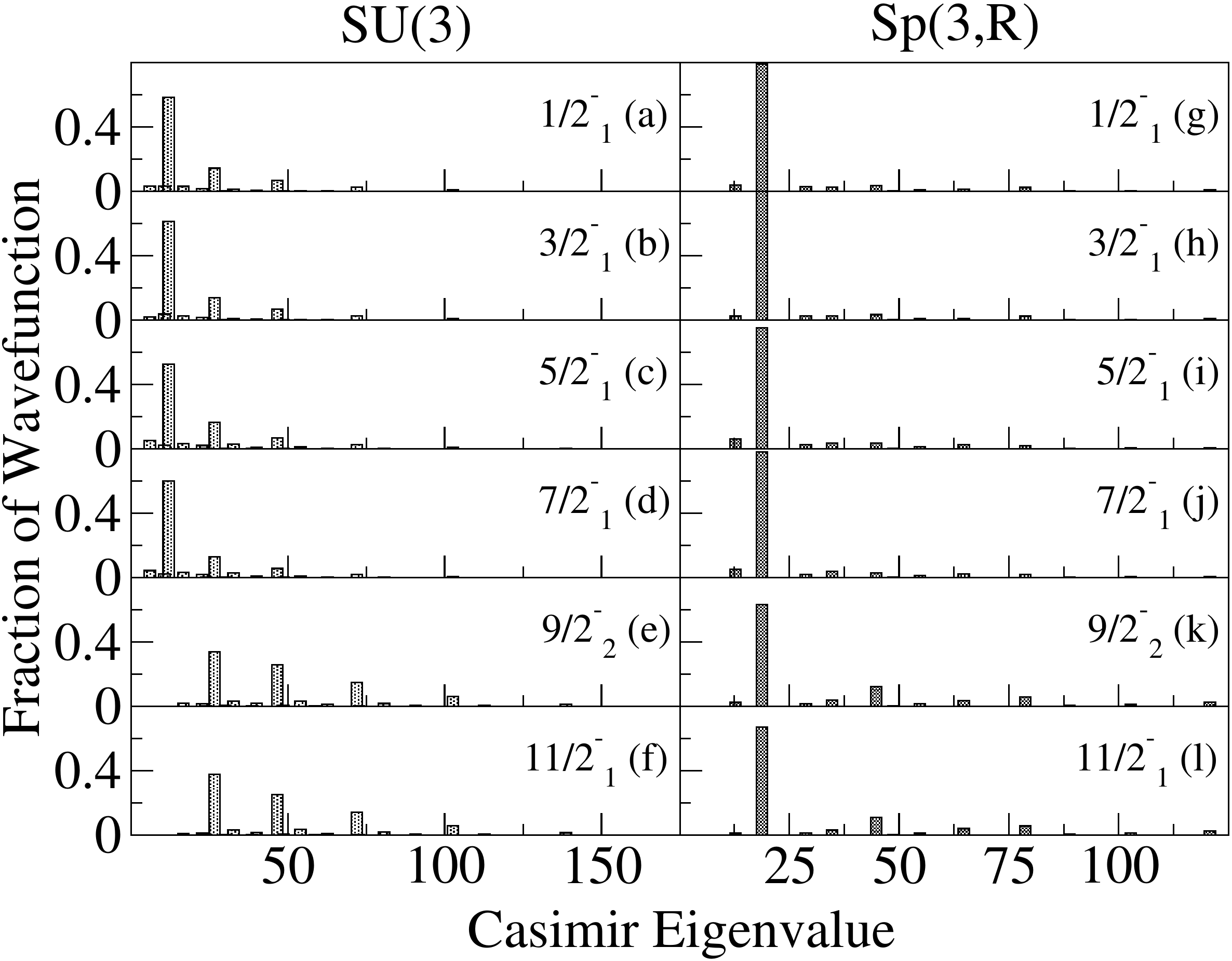}

\caption{Algebraic decompositions of the ground-state band of $^7$Be shown in
  Fig.~\ref{fig:be7be2}. Left-hand panels (a)-(f): decomposition by the
  quadratic Casimir of $\mathrm{SU}(3)$. Right-hand panels (g)-(l):
  decomposition by the quadratic Casimir of $\mathrm{Sp}(3,R)$. The maximal
  angular momentum within the lowest harmonic oscillator configuration, the
  so-called valence subspace, is $J=7/2$.}
\label{fig:be7decomp_a}
\end{figure}

Now we come to the main motivation for this work. The Elliott $\mathrm{SU}(3)$
model has a long and successful history of describing rotational bands. Yet it
has a natural limitation: $\mathrm{SU}(3)$ irreps are constrained to configurations involving the valence
shell, or, more generally, they cannot combine configurations from different
shell model $\Nex \hbar \omega$ spaces.  If one expects low-lying states to be
dominated by valence subspace configurations, and the valence subspace has a
maximum angular momentum, then the rotational band should terminate at that
maximum angular momentum.

A long-standing question has been of extended bands,
that is, bands that contain members outside the valence subspace.  An early
phenomenological study of $^8$Be and $^{20}$Ne based upon particle-hole
excitations (and not in an algebraic picture) found only weak $E2$ transitions and
thus no extended band \cite{goode1970ground}.  Later phenomenological [but using
an $\mathrm{Sp}(3,R)$ framework] calculations of
$^{20}$Ne~\cite{ap-126-1980-343-Rosensteel,draayer1984symplectic}, as well as
\emph{ab initio} NCSM calculations of beryllium
isotopes~\cite{PhysRevC.91.014310}, found evidence for extended rotational
bands: strong transitions from states outside the valence subspace, and in
particular beyond the maximally allowed valence-subspace angular momentum, to
ground-state band members.

While this violates the Elliott $\mathrm{SU}(3)$ picture, note that the
physical electric quadrupole ($E2$) transition operator~\cite{BG77},
proportional to $\sum_i e_i r_i^2 Y_{2m}(\hat{r}_i)$ where $e_i$ is the 
charge of the $i$ nucleon, connects configurations of different $N_A$, while the
Elliott quadrupole operator~(\ref{ElliottQ}), cannot.   Because the isoscalar component 
of the $E2$ transition  operator is encompassed in $\mathrm{Sp}(3,R)$, 
we choose to carry out a side-by-side comparison 
of $\mathrm{SU}(3)$ and $\mathrm{Sp}(3,R)$ symmetry for such bands.
Specifically, we take NCSM calculations of several beryllium isotopes, as well
as of $^{20}$Ne, and, using previously developed techniques, decompose the wave
functions into the irreps of Elliott's $\mathrm{SU}(3)$ and of
$\mathrm{Sp}(3,R)$.  Note that the NCSM calculations make no assumptions
about the underlying group structure.  

One can adapt 
the Lanczos algorithm
 to decompose any wave function into its contributions from
 subspaces defined by the eigenvalues of any Hermitian operator~\cite{whitehead1980:lanczos}.
Thus, to find $\grpsu{3}$ or $\grpsptr$ decompositions of an NCSM wave functions obtained in an
ordinary $M$-scheme basis, we make use of the Casimir operator for the given
group.
Lanczos-enabled decompositions have been used before
to investigate quasi-dynamical symmetry~\cite{PhysRevC.63.014318,PhysRevC.91.034313,PhysRevC.95.024303}.

A \textit{Casimir operator} $\hat{C}$ is  built from the generators of a given group, 
constructed so as to
commute with all the generators (equivalently, it is invariant under the
symmetry operations of the group).  It therefore  acts as a constant
multiple of the identity operator within any irrep, and has an eigenvalue
which  depends only upon the quantum numbers labeling the irrep:
$\hat{C}|\alpha \Gamma i \rangle = g(\Gamma)|\alpha \Gamma i\rangle$.
For the familiar case of $\grpsu{2}$, the quadratic
Casimir operator is $\hat{C}_2[\grpsu{2}]=\hat{\vec{J}}\cdot\hat{\vec{J}}$, with eigenvalue $g(J)=J(J+1)$.
The quadratic Casimir operator of $\grpsu{3}$ is, to within an arbitrary choice of normalization~\cite{harvey1968nuclear,rowe1996dynamical,PhysRevC.65.054309},
\begin{equation}
\hat{C}_{2}[\grpsu{3}] = \frac{1}{6}\left[\hat{\mathcal{Q}}\cdot\hat{\mathcal{Q}} +
  3\hat{\vec{L}}\cdot\hat{\vec{L}}\right],
\label{su3casimir}
\end{equation}
which has eigenvalue
\begin{equation}
g(\lambda,\mu)=\frac23(\lambda^{2} + \lambda\mu + \mu^{2} + 3\lambda + 3\mu).
\end{equation}
The quadratic $\grpsptr$ Casimir
operator for $\grpsptr$ is~\cite{PhysRevC.65.054309}
\begin{equation}
    \hat{C}_2[\grpsptr]=\hat{C}_2[\grpsu{3}]-2\hat{A}_{2}\cdot
    \hat{B}_2-2\hat{A}_{0}\cdot \hat{B}_0+\frac13\hat{H}_0^2-4\hat{H}_0,
\end{equation}
with eigenvalue
\begin{equation}
g(\sigma)=\frac23(\lambda_\sigma^2+\mu_\sigma^2+\lambda_\sigma\mu_\sigma+3\lambda_\sigma+3\mu_\sigma)+\frac{N_\sigma^2}{3}-4N_\sigma.
\end{equation}

The basic idea, then, is that the full nuclear many-body space can be broken into
subspaces identified with an eigenvalue of the Casimir operator.  This normally
corresponds to selecting a particular value for the irrep label $\Gamma$, though
there can be ambiguities: \textit{e.g.}, for $\mathrm{SU}(3)$, irreps with
labels $(\lambda, \mu)$ or $(\mu,\lambda)$ have the same eigenvalue for the
quadratic Casimir operator~(\ref{su3casimir}).  [One can, in principle, separate
  out contributions from these irreps by using higher-order Casimir operators,
  such as the cubic Casimir for $\mathrm{SU}(3)$, which break the degeneracy,
  but this involves going to three-body or higher-body operators in the shell
  model calculation.]

One can adapt the Lanczos algorithm to compute the
fraction $\mathcal{F}(\Gamma)$ of a given
wave function, in any such subspace $\Gamma$~\cite{PhysRevC.63.014318,PhysRevC.91.034313,PhysRevC.95.024303}.
This is equivalent to projecting the state $|\Psi\rangle$ onto a full group theoretical basis:
\begin{equation}
\mathcal{F}(\Gamma) = \sum_{\alpha,i }| \langle \alpha \Gamma i| \Psi \rangle|^{2}.
\end{equation}
For instance, for $\grpsu{3}$, in finding the fraction from $\Gamma\rightarrow
(\lambda,\mu)$, we implicitly aggregate all contributions from the many ways
this $(\lambda,\mu)$ can be obtained as the result of different shell model
configurations and intermediate couplings, as well as the contributions from
different states $i$ within each irrep.  Luckily, the Lanczos method efficiently
accomplishes this without any need to explicitly construct a group
theoretical
basis.

In this paper we provide decompositions by plotting $\mathcal{F}(\Gamma)$, the
fraction of wavefunction in the subspace, versus $g(\Gamma)$.
An example is in Fig.~\ref{fig:be7decomp_a}.  As we follow members of the ground
state band, defined by strong $E2$ transitions, to higher angular momentum, we
find, as expected, that the Elliott $\mathrm{SU}(3)$ decompositions change
abruptly as one exits the valence subspace, that is, as the band extends to angular
momenta not contained in the valence subspace.  Decompositions by the
$\mathrm{Sp}(3,R)$ quadratic Casimir, on the other hand, are nearly identical
across the valence subspace boundary, providing a more consistent representation of
the members of a band than those by the Elliott $\mathrm{SU}(3)$ Casimir. This
is the overarching theme of this paper.

\begin{figure}[h]
\centering
	\includegraphics[width=0.9\linewidth]{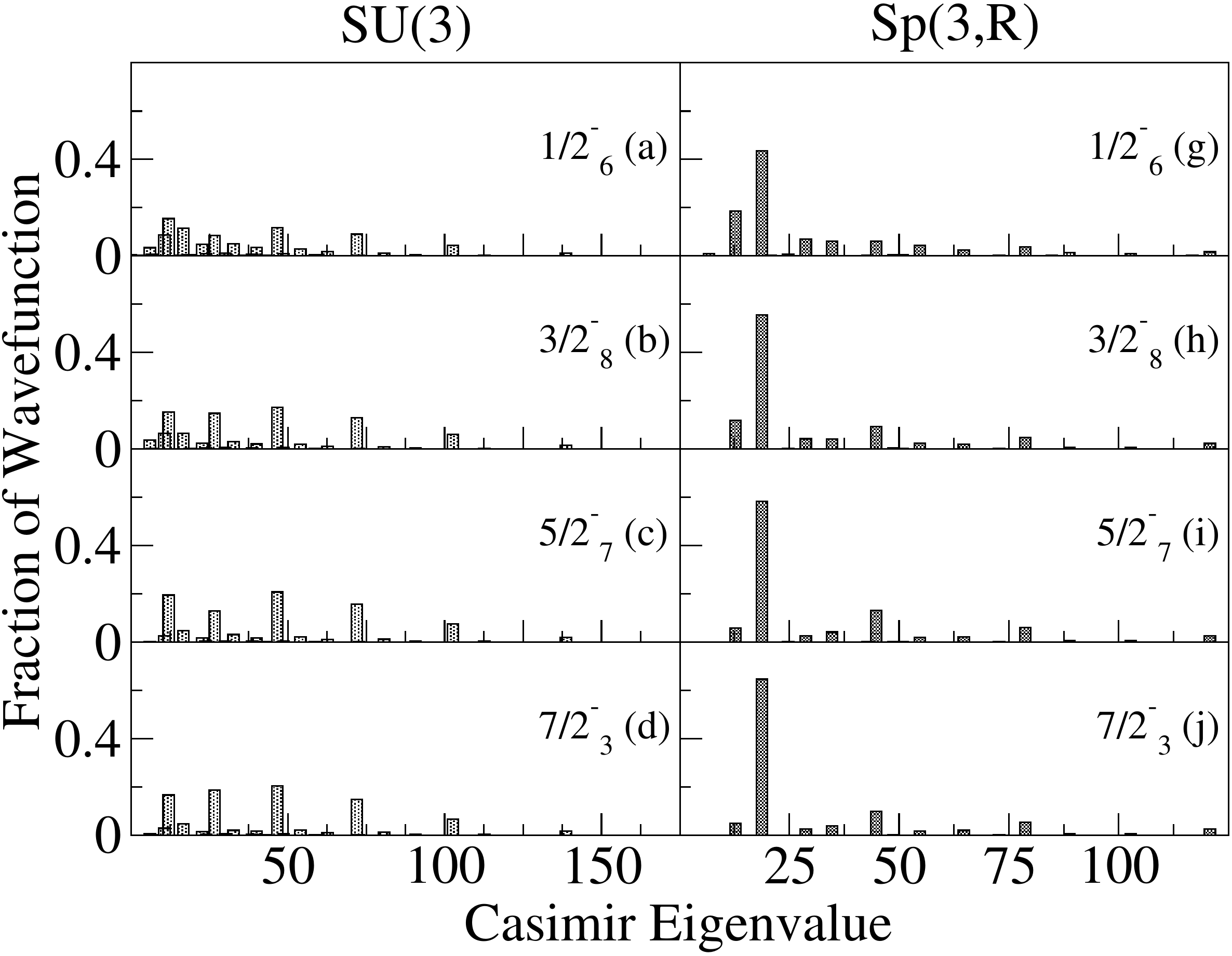} 
	\caption{Decompositions of members of the upper rotational band of
          $^7$Be shown in Fig.~\ref{fig:be7be2}. Left-hand panels, (a)-(f):
          Decomposition by the quadratic Casimir of $\mathrm{SU}(3)$. Right-hand
          panels,(g)-(l): decomposition by the quadratic Casimir of
          $\mathrm{Sp}(3,R)$. Although not shown, these states are all dominated
          by $\Nex=2$ ($2\hbar \omega$) configurations. }
	\label{fig:be7decomp_b}
\end{figure}

Furthermore, as can be seen in Fig.~\ref{fig:be7be2}, the $E2$ transitions
connect not only the yrast band, but also extends to an upper band consisting of
excited states connected by strong $B(E2)$ values.  In
Fig.~\ref{fig:be7decomp_b} we decompose these states: these states share the
same $\mathrm{Sp}(3,R)$ decomposition, but a very different (albeit consistent)
$\mathrm{SU}(3)$ decomposition from the ground-state band
\cite{mccoy2018symplectic,mccoy2020emergent}. We show this pattern persists not
only in several beryllium isotopes, but also in $^{20}$Ne.  Because the two bands 
unite at high angular momentum and share the same symplectic decomposition, we  argue they are really a single unified feature rather than independent bands.

\section{Results}

In this section we discuss in some detail our results, working through each of the studied nuclides one by one.

\medskip

\begin{figure}[h]
\centering
\begin{tabular}{c}
	\includegraphics[width=0.9\linewidth]{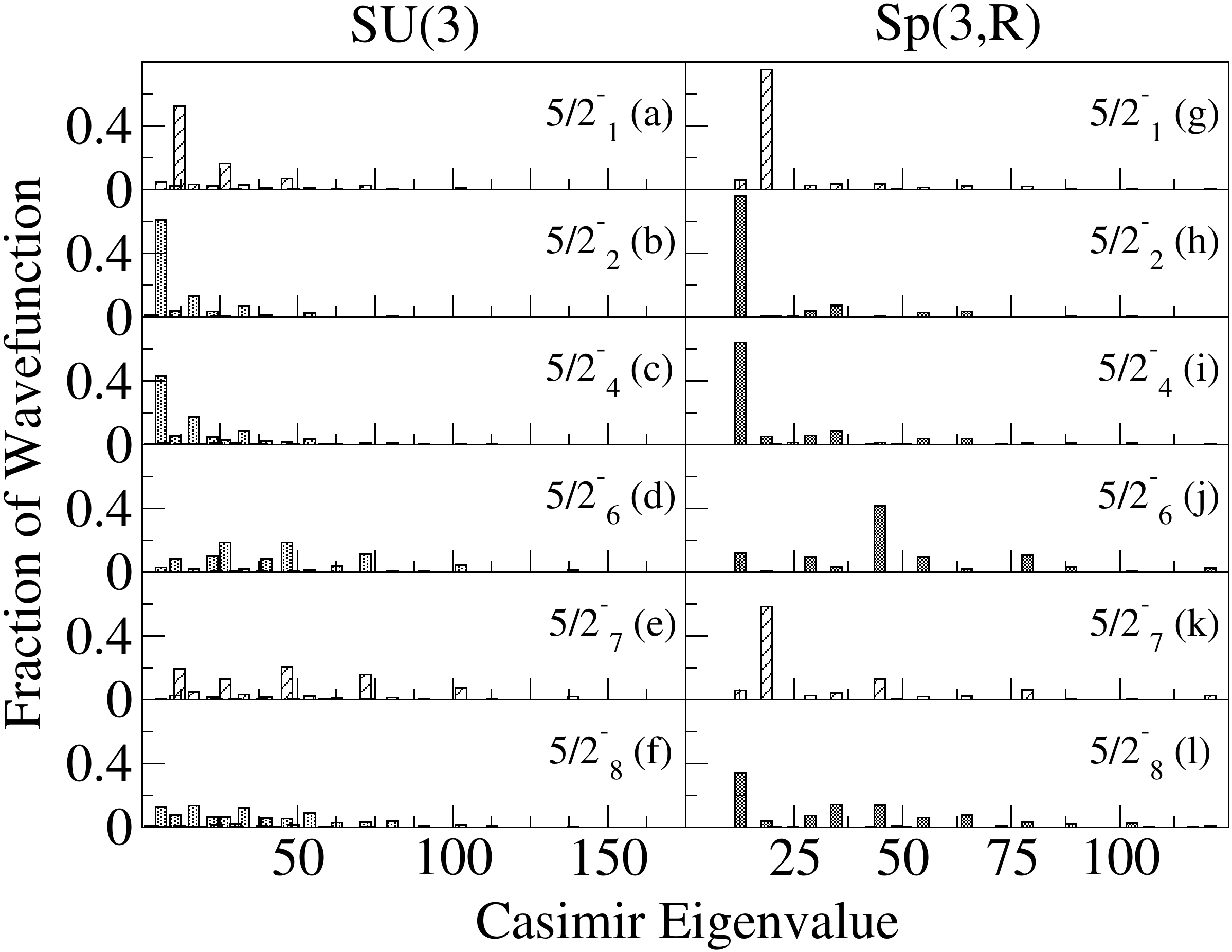}
\end{tabular}
	\caption{Decompositions of $5/2^-$ states of ${}^{7}$Be.
	Left-hand panels (a)-(f): decomposition by the quadratic Casimir of $\mathrm{SU}(3)$. Right-hand panels (g)-(l): decomposition by the quadratic Casimir of $\mathrm{Sp}(3,R)$. Decompositions duplicated for purposes of 
	comparison from Fig.~\ref{fig:be7be2}, 
	that is, the $5/2^-_1$ and $5/2^-_7$ states, are marked with stripes.  }
	\label{fig:be7decomp_c}
\end{figure}

\subsection{$\isotope[7]{Be}$}
Previous works~\cite{caprio2013emergence,PhysRevC.91.014310,caprio2015collective,caprio2020probing,mccoy2020emergent} 
have made detailed 
investigations into $^7$Be as a rotational nucleus in \textit{ab initio} calculations, looking carefully at  ratios of 
$B(E2)$ values, $M1$ transitions, decomposition of the wavefunction into total orbital 
angular momentum $L$ and total spin $S$, and so on.   
Those calculations were made using the JISP16 \cite{JISP16}, NNLO\textsubscript{opt}~\cite{ekstroem2013:nnlo-opt},  and Daejeon16 \cite{Daejeon16} interactions. 
While we follow in their footsteps, we use an interaction derived from an 
N3LO chiral effective field theory \cite{machleidt2011chiral}, and we simply identify band structure through 
strong $B(E2)$ values.

Fig.~\ref{fig:be7be2} shows both our calculated excitation spectrum 
(with known experimental levels \cite{TILLEY20023} marked for comparison) and, through strong $B(E2)$s, clear band structure. 
The ground-state band of  $^7$Be has the Coriolis staggering typical of  $K=1/2$ bands, where $K$ is 
the $z$-projection of angular momentum in the intrinsic frame \cite{ring2004nuclear}, with an inverted angular momentum sequence ($3/2^-_1$, $1/2^-_1$, $7/2^-_1$, $5/2^-_1$). 
Later on we will look at another $K=1/2$ band in the unnatural parity space of $^9$Be.

We plot the downward $B(E2)$ strengths as solid lines, with the width and
shading approximately proportional to the log strength, for $B(E2)$ values above
1.5 $e^2\mathrm{fm}^4$. It is also important to note in Fig.~\ref{fig:be7be2}
many downward transitions drawn between states are not part of the identified
bands. This is done to illustrate how we use $E2$ transition network plots as a qualitative tool
to effectively locate theoretical rotational structure amongst many other
transitions. However, for nuclei in the following sections, we omit such $E2$
transitions not involving states we have identified as members of rotational
bands.

As discussed in Sec.~\ref{sm}, a key concept  is the `valence subspace'
 defined by the lowest harmonic oscillator configurations, that is, $\Nex=0$ or $0\hbar\omega$ configurations.  
  For $^7$Be and the other beryllium isotopes we consider, the valence subspace consists of
 an inert $^4$He core and the remaining nucleons restricted to the $0p$ shell. In particular we are 
 interested in the maximal angular momentum in the valence subspace, which is $J=7/2$ for $^7$Be.
Hence any state with $J > 7/2$ must be outside the valence subspace. In  Fig.~\ref{fig:be7be2}, and  subsequent similar figures, we denote the maximal $J$ in the valence space by a vertical dashed line. 
 
Given the limit in angular momentum in the valence subspace, it is unsurprising
that the $\mathrm{SU}(3)$ decompositions shown in Fig.~\ref{fig:be7decomp_a}
(a)-(f) express an abrupt change in dominant irreps as one goes from the
$7/2_1^-$ state to the $9/2^-_2$ state.  On the basis of Elliott's
$\mathrm{SU}(3)$ model, then, one would naively expect that the ground-state
band should terminate at $J=7/2$. But the $B(E2)$ values show strong transition
strengths to both the $9/2^-_2$ and $11/2^-_1$ states, and hence by that
traditional criteria these states are candidates for belonging to the same band.
Yet $9/2^-_2$ and $11/2^-_1$ states of the lowest band are, of necessity, outside the valence subspace, 
and  are dominated by $\Nex=2$ configurations \cite{caprio2020probing}.  We remind the reader 
that within a single $\mathrm{Sp}(3,R)$ irrep one can have states belonging to different $\mathrm{SU}(3)$ irreps, involving different numbers $\Nex$ of oscillator excitations.

It is very satisfactory, then, to see this analysis borne out when decomposing in
$\mathrm{Sp}(3,R)$, in the right-hand panels Fig.~\ref{fig:be7decomp_a}(g)-(l). All of the
states in the band show nearly identical decompositions, and indeed are close to
a pure dynamical symmetry. This provides evidence $\mathrm{Sp}(3,R)$ is a more unifying
symmetry for rotational bands than Elliott's $\mathrm{SU}(3)$.  Of course, this was the
original motivation for $\mathrm{Sp}(3,R)$ \cite{PhysRevLett.38.10}; as in
references~\cite{mccoy2018:diss,mccoy2020emergent}, we see this relationship emerge without any
assumptions, starting from an independent \textit{ab initio} nuclear interaction
and relying upon $B(E2)$ values to identify bands.

The ground-state band does not exist in isolation: there is an upper band~\cite{mccoy2020emergent}, 
connected by strong $B(E2)$s, including  the 
 $1/2^-_6$, $3/2^-_8$, $5/2^-_7$ and $7/2^-_3$ as shown in Fig.~\ref{fig:be7be2}.
 (This is for our $N_\mathrm{max} = 10$ calculations. Because of the high density of excited states, the exact position of these members of the excited band 
 will depend sensitively upon the details of the calculation, not only the model space and 
 single-particle basis frequency, but also upon the choice of interaction.
 Specifically, we do not claim, for example, that the physical $3/2^-_8$ state is 
 a member of the band, only that there exists a highly excited $3/2^-$ state which is a member of the band.
 Our broader narrative, however, 
 that is the \textit{existence} of a strongly connected upper band, is robust and insensitive to these details.) 
 These two `bands', as traced out by $E2$ transitions, appear to share members at high angular momentum, here $J=11/2$, a pattern we will see repeated in other 
 nuclei.
 
We decompose the  states identified as members of the upper band in Fig.~\ref{fig:be7decomp_b}.
The $\mathrm{SU}(3)$ decompositions for the $1/2_6^-, 3/2^-_8, 5/2^-_7$, and $7/2^-_3$ states,  Fig.~\ref{fig:be7decomp_b}(a)-(d), while highly fragmented, are nonetheless similar to each other, 
but very different from the $\mathrm{SU}(3)$ decompositions of the ground-state band.  
The decompositions by $\mathrm{Sp}(3,R)$,   Fig.~\ref{fig:be7decomp_b}(g)-(l), tell a different story. A distinct dominant $\mathrm{Sp}(3,R)$ irrep is shared across all the states, and despite some fragmentation in the $1/2_6^-$ and $3/2_8^-$ in   Fig.~\ref{fig:be7decomp_b}(g), (h), which we 
attributed to mixing due to the high local density of states with similar $J^\pi$,
the upper band decompositions directly resemble the ground-state band in Fig~\ref{fig:be7decomp_a}.

To show that $\mathrm{Sp}(3,R)$ differentiates band members from states outside the band,
we decompose several $5/2^-$ states in Fig.~\ref{fig:be7decomp_c}.  The $5/2_1^-$
and $5/2_7^-$ states have already been discussed as members of the ground-state band and
upper band, respectively.  Recall that the states in these bands have similar $\mathrm{Sp}(3,R)$
decompositions [as seen in Fig.~\ref{fig:be7decomp_c} (g), (k)] even though the $\mathrm{SU}(3)$
decompositions have little in common [as seen in Fig.~\ref{fig:be7decomp_c} (a), (e)].
The other $5/2^-$ states, not members of either band, have very different $\mathrm{Sp}(3,R)$ decompositions.
We have carried out other similar studies, \textit{i.e.}, decomposed multiple excited states of the same angular momentum and parity with the same result: 
states outside the band have dramatically different $\mathrm{Sp}(3,R)$ decompositions.

Our calculations also naturally 
 determine the fraction of the wave function for each $\Nex$ (or $N_\mathrm{ex}
 \hbar \omega$).  As prior work~\cite{caprio2020probing} discusses the
 distribution in $\Nex$ in more detail, we only touch upon the highlights.
 As we increase $N_\mathrm{max}$, the fraction of the wave function in the valence subspace naturally decreases.
  For example, at $N_\mathrm{max} =
 2$ the yrast band states have $\sim 85\%$ of their probability coming from $\Nex=0$ configurations, while at
 $N_\mathrm{max} = 6$ this contribution falls to $\sim 71\%$. By
 $N_\mathrm{max}=10$, where the distribution appears to stabilize, yrast states
 with $J \leq 7/2$ still had $\sim 60\%$ of their wave function within the
 valence subspace.
 
 By contrast, the highly excited $5/2^-_7$ state has roughly $20\%$ of its wavefunction in each of the $\Nex = 0$ and $\Nex = 2$ spaces and $26\%$ in the $\Nex = 4$ space.
 Correspondingly, the $5/2^-_7$ has a very different SU(3) decomposition from the yrast $5/2^-_1$
 which has the largest fraction of its wave function in the valence subspace.
 Yet, as we have already noted, the $5/2^-_1$ and $5/2^-_7$
share nearly identical 
 $\mathrm{Sp}(3,R)$ decompositions.

This trend continues: the $5/2^-_2$ and $5/2^-_4$ states, like the $5/2^-_1$ state, have wave 
functions with a largest fraction in the valence subspace, and have similar
decompositions in $\mathrm{SU}(3)$, Fig.~\ref{fig:be7decomp_c} (b), (c) and $\mathrm{Sp}(3,R)$,
Fig.~\ref{fig:be7decomp_c} (h), (i).
Conversely, $5/2^-_6$ is the first $5/2^-$ state to be primarily outside the valence subspace, 
with  only a few percent contribution from $\Nex=0$, and   $\sim 30\%$ from $\Nex=2$  
and $\Nex=4$ configurations each. Correspondingly one can see the decompositions in both $\mathrm{SU}(3)$, 
in Fig.~\ref{fig:be7decomp_c} (d), and $\mathrm{Sp}(3,R)$, in Fig.~\ref{fig:be7decomp_c} (j), are very different 
from the lower-lying $5/2^-$ states. 

The above calculation was carried out at $N_\mathrm{max} = 10$. As
$N_\mathrm{max}$ increases, although the ground-state energy (or binding energy)
evolves significantly, excitation energies of the states which lie primarily in
the valence subspace converge very quickly (see, \textit{e.g.}, figure~6 of
reference~\cite{caprio2020probing} for illustration). The excitation energies of
the states involving primarily contributions from higher $\Nex$ converge more slowly (figure~17 of
reference~\cite{caprio2020probing}).  The
decompositions, however, are generally robust with respect to $N_\mathrm{max}$
and, to a lesser extent, the basis oscillator frequency, as detailed in the Appendix
(figures~\ref{fig:be7vNmax} and \ref{fig:be7vHw}).

\begin{figure}[h]
\centering
	 \includegraphics[width=0.9\linewidth]{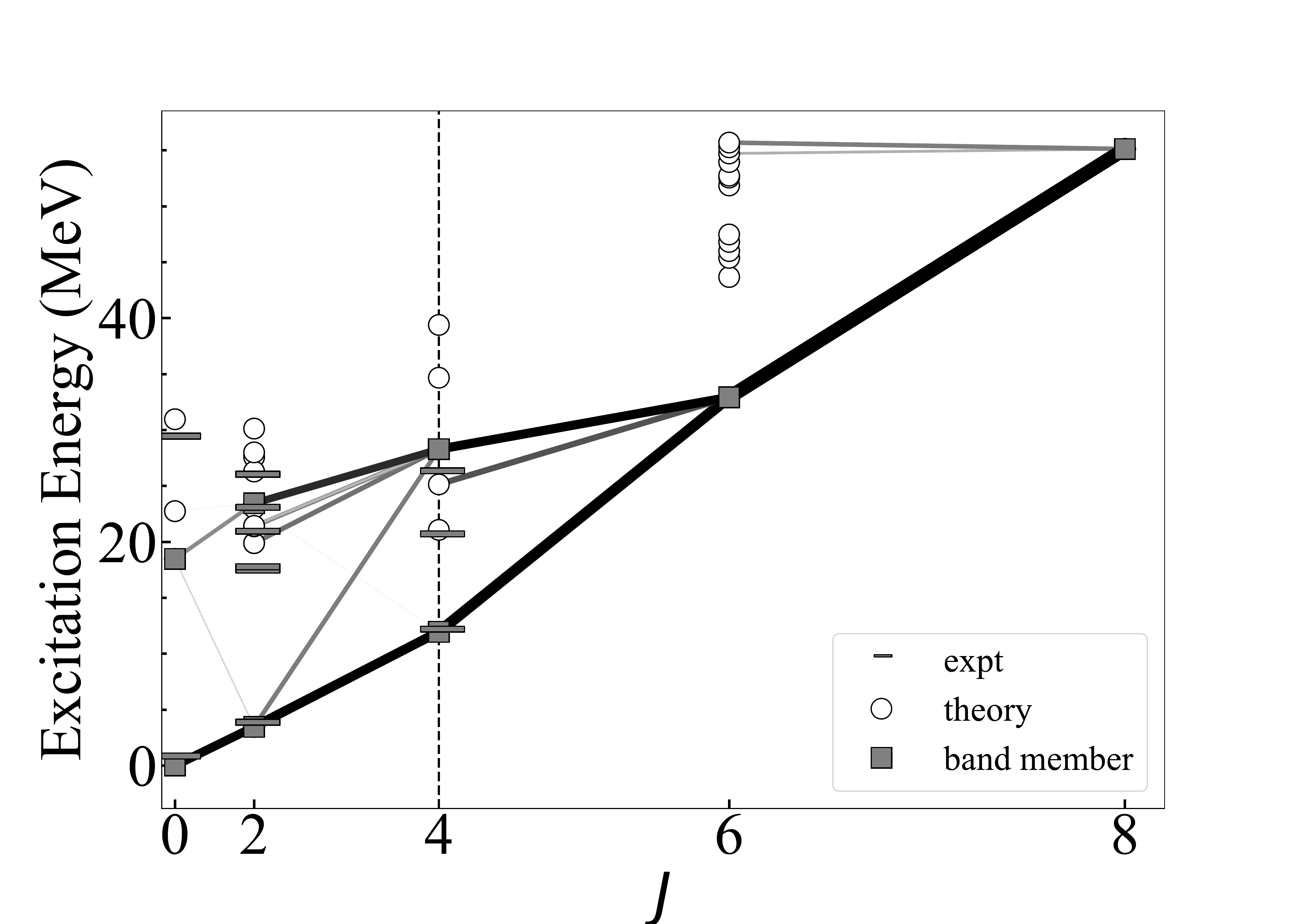} 
	 \caption{Excitation spectrum of the natural positive parity states of
           ${}^{8}$Be, computed at $N_\mathrm{max} =10$ with a single-particle
           harmonic oscillator basis frequency $\hbar \omega = 20$ MeV, using a
           two-body chiral force at N3LO. Experimental energies \cite{TILLEY2004155} are presented as
           shaded bars for comparison (see text). Shaded squares indicate  states  decomposed into group irreps
           in Fig.~\ref{fig:Be8decomp_a}. Downward (high to low $J$) $E2$
           transitions connecting to band members are indicated with solid 
           lines. Line thickness and shading is proportional to $B(E2)$
           strength. Only strengths above 0.1 $e^2 \mathrm{fm}^4$ are shown. The
           vertical dashed line denotes the maximal angular momentum within the
           lowest harmonic oscillator configuration.}
	\label{fig:Be8Be2}
\end{figure}

 \begin{figure}
 \centering
\begin{tabular}{c}
	\includegraphics[width=0.7\linewidth]{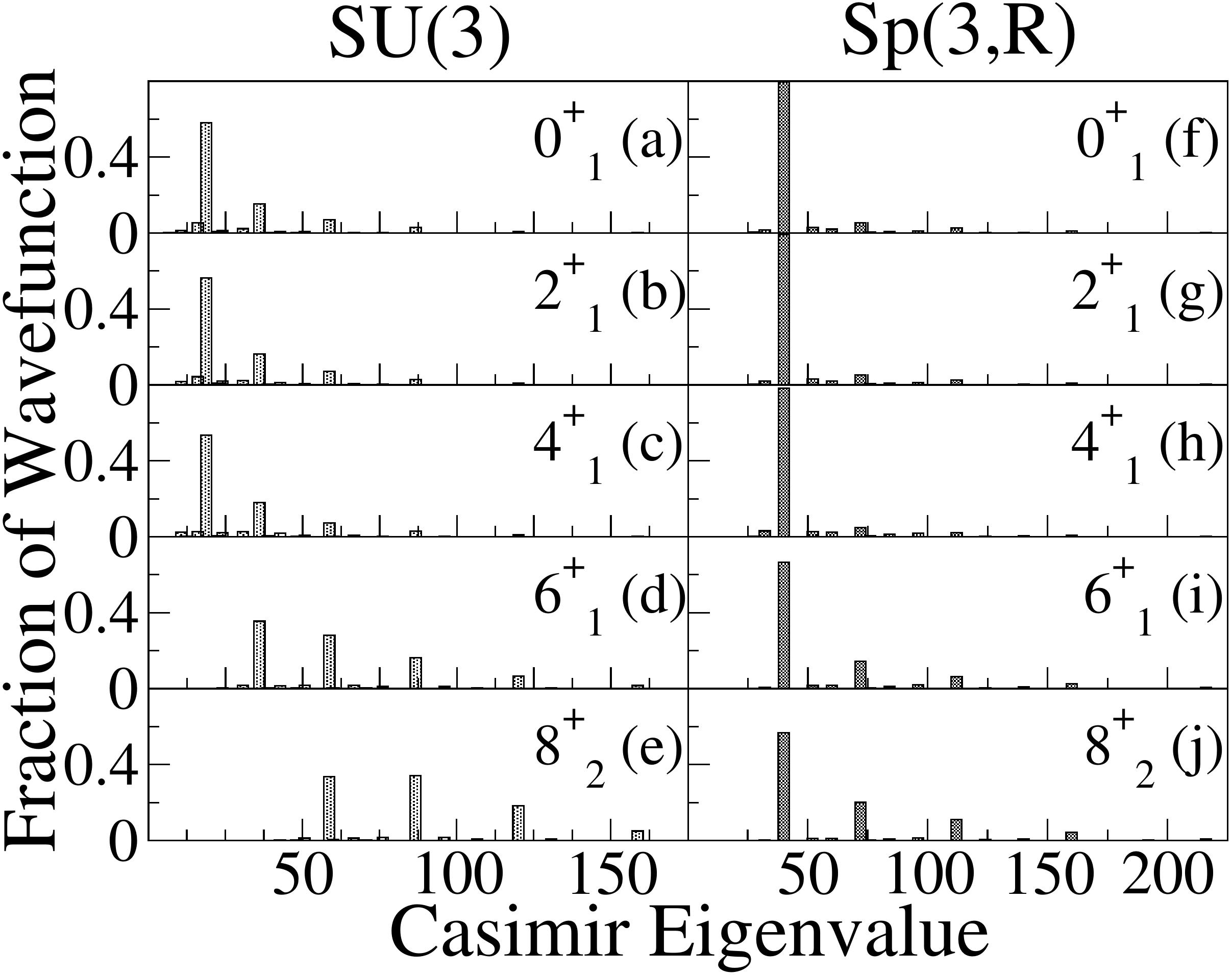}
 \\
	\includegraphics[width=0.7\linewidth]{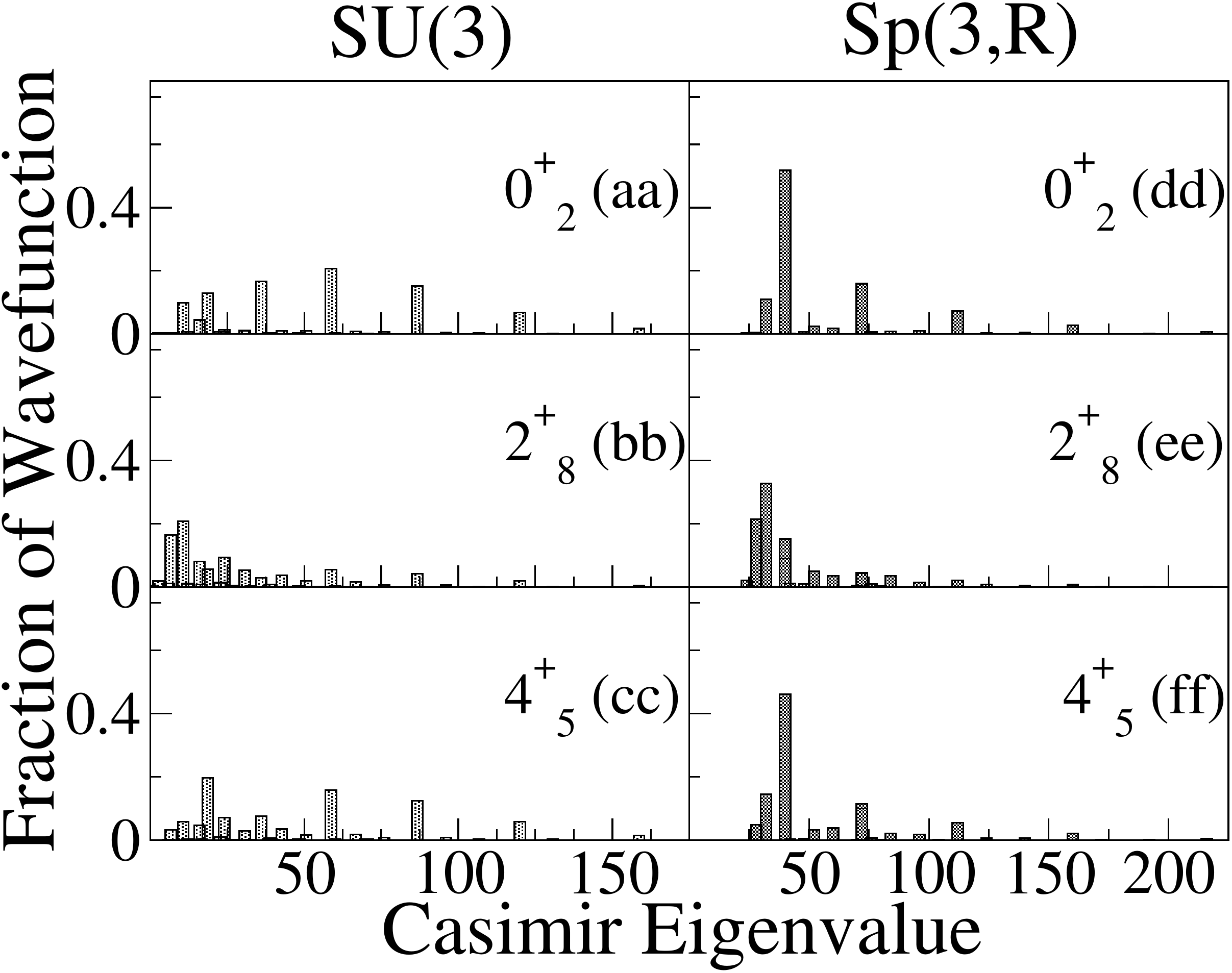}
		\end{tabular}

	\caption{Decompositions by the quadratic Casimir of $\mathrm{SU}(3)$ and $\mathrm{Sp}(3,R)$ of $^8$Be band members as calculated in Fig.~\ref{fig:Be8Be2}. Ground-state decompositions are shown in panels (a)-(j). Excited band decompositions are shown in panels (aa)-(ff).}
	\label{fig:Be8decomp_a}
	
\end{figure}

We consider three more nuclides to demonstrate these patterns are not a singular phenomenon. 

\subsection{$\isotope[8]{Be}$} 

The strongly-connected positive parity $K=0$ \cite{caprio2015collective} yrast band for
$^8$Be, seen in Fig.~\ref{fig:Be8Be2}, is not surprising, given that one can
think of $^8$Be as two $\alpha$ particles in a dumbbell configuration
\cite{yoshida2013cluster}. Nonetheless, the story here echoes that of $^7$Be.

The maximum angular momentum in the valence subspace is $J=4$, but our calculated $B(E2)$s clearly
demarcate a band extending out of the valence subspace to the $6^+_1$ and $8^+_2$
states. 
Investigation of a possible extension of the ground-state band in
$^8$Be to extra-valence states go back at least fifty years.  While a 
two-particle, two-hole calculation of 
extra-valence states found only weak $B(E2)$ values, suggesting there was not an extended
band~\cite{goode1970ground}, other calculations supported an extended band. 
These include  projected Hartree-Fock calculations~\cite{bouten1970ground} 
and configuration-interaction calculations 
investigating the mixing of $\mathrm{SU}(3)$~\cite{bouten1971ground}. In particular early multi-$\Nex$ calculations
\cite{arickx1975configuration}  found, like us,  lower and upper
bands; this  led to the use of $\mathrm{Sp}(2,R)$ \cite{arickx1976new,arickx1979sp} and 
prefigured later use of $\mathrm{Sp}(3,R)$.  Finally, more recent
investigations in NCSM frameworks likewise found
the $6^+_1$ and $8^+_2$ states to be members of the ground-state band~\cite{PhysRevC.91.014310,mccoy2018:diss},  
and identified an upper band 
 having very similar $\mathrm{Sp}(3,R)$ decompositions to the ground-state band~\cite{mccoy2018:diss}.

Fig.~\ref{fig:Be8decomp_a}(a)-(j) (left-hand
plots) clearly shows
that decompositions with $\mathrm{Sp}(3,R)$ provide a more unifying picture than
with $\mathrm{SU}(3)$. Decompositions with Elliott's $\mathrm{SU}(3)$ are nearly identical
for the yrast states within
the valence subspace, but change dramatically for $J > 4$ states  outside
the valence subspace.  There is some evolution of the $\mathrm{Sp}(3,R)$
decomposition also for $J > 4$ ground-state band members, but much less dramatic.
Note that while we went up to $N_\mathrm{max}=10$, the rate of convergence with increasing
$N_\mathrm{max}$ for the energies of the extra-valence $6^+_{1}$, and $8^+_2$
states is significantly slower than for states with $J\leq4$, much as we found
with other nuclides.

Strong $E2$ transitions clearly link the upper band, $0^+_{2}$, $2^+_{8}$,
$4^+_{5}$ at $N_\mathrm{max}=10$, to the ground-state band.  
Fig.~\ref{fig:Be8decomp_a} (aa)-(ff) (right-hand plots) presents decompositions of
this upper band.  The $\mathrm{Sp}(3,R)$ decompositions of the upper band
members in Fig.~\ref{fig:Be8decomp_a}(dd)-(ff) are very similar to those of the
ground-state band, albeit with some fragmentation. The fragmentation is especially noticeable
for the $2_8^+$ state, which could be due to mixing with nearby states, while the
$\mathrm{SU}(3)$ decompositions in Fig.~\ref{fig:Be8decomp_a}(aa)-(cc) are
highly fragmented.  We note the valence
yrast ($0^+_1$, $2^+_1$ and $4^+_1$) wave functions  have $\Nex=0$ contributions
of roughly $60\%$, similar to that of $^7$Be, while members of the upper band
  with $J \leq 4$ had a smaller fraction of their wave functions in the
  $N_\mathrm{ex}=0$ space, between $20-45\%$.  This follows prior work using a symplectic
  $\Nmax=4$ basis
that found  the upper band states and the $6^+_1$ and $8_2^+$ states have  
similar $U(3)$ decompositions~\cite{mccoy2018:diss}.

\begin{figure}[h]
\centering
	\includegraphics[width=0.9\linewidth]{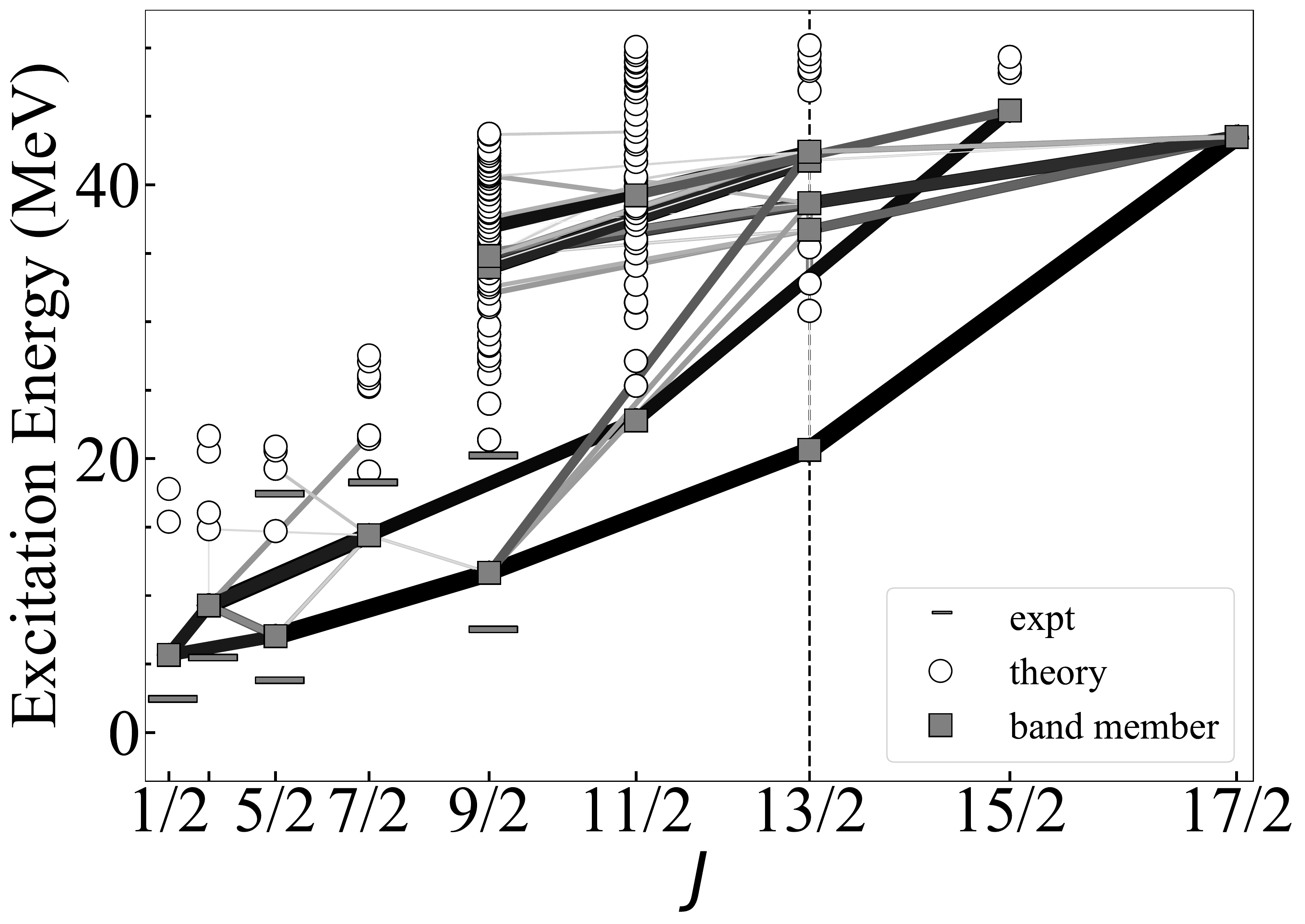}
\caption{Excitation spectrum of the unnatural, positive parity states of
  ${}^{9}$Be, computed at $N_\mathrm{max} =9$ with a single-particle harmonic
  oscillator basis frequency $\hbar \omega = 20$ MeV, using a two-body chiral
  force at N3LO. (The natural parity states are not shown.) Excitation energies
  are calculated with respect to the natural (negative) parity ground state. Experimental
  energies are presented as shaded bars for comparison (see text).  
  Shaded squares indicate states we decompose 
  into group irreps in Fig.~\ref{fig:Be9decomp_a} and \ref{fig:Be9decomp_b}. 
  $J$-changing $E2$ transitions connecting to band members are indicated with solid 
  lines. Line thickness and shading is proportional to $B(E2)$ strength. Only
  strengths above 2.0 $e^2\mathrm{fm}^4$ are shown. The vertical dashed line
  denotes the maximum angular momentum in
  the $\Nex =1$ space. }
	\label{fig:Be9Be2}
\end{figure}

\begin{figure}[h]
\centering
	\includegraphics[width=0.9\linewidth]{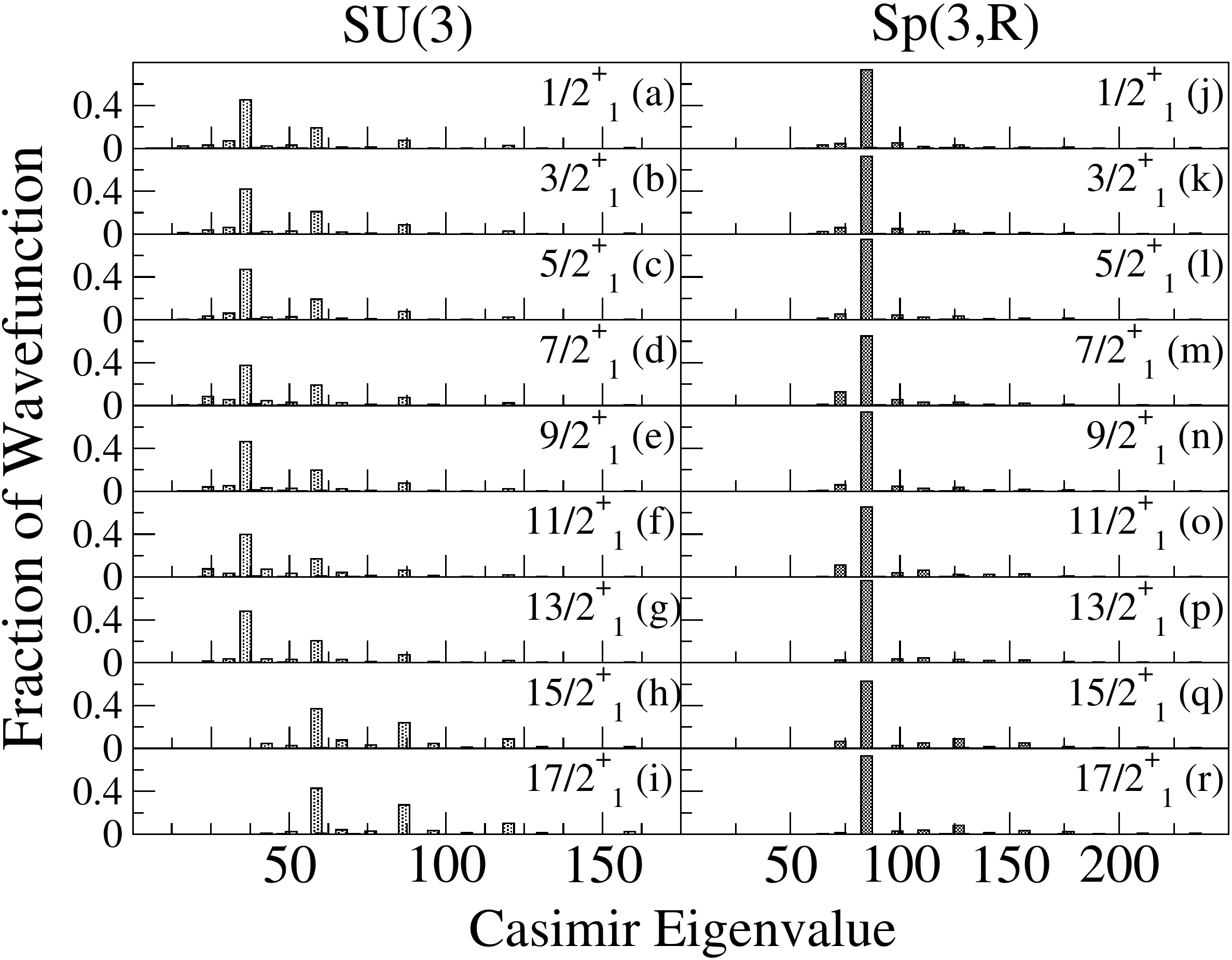}
	\caption{Decompositions of the members of the lower rotational band of
          $^9$Be shown in Fig.~\ref{fig:Be9Be2}. Left-hand panels, (a)-(i):
          Decomposition by the quadratic Casimir of $\mathrm{SU}(3)$. Right-hand
          panels,(j)-(r): decomposition by the quadratic Casimir of
          $\mathrm{Sp}(3,R)$.}
	\label{fig:Be9decomp_a}
\end{figure}

\begin{figure}[h]
\centering
	\includegraphics[width=0.9\linewidth]{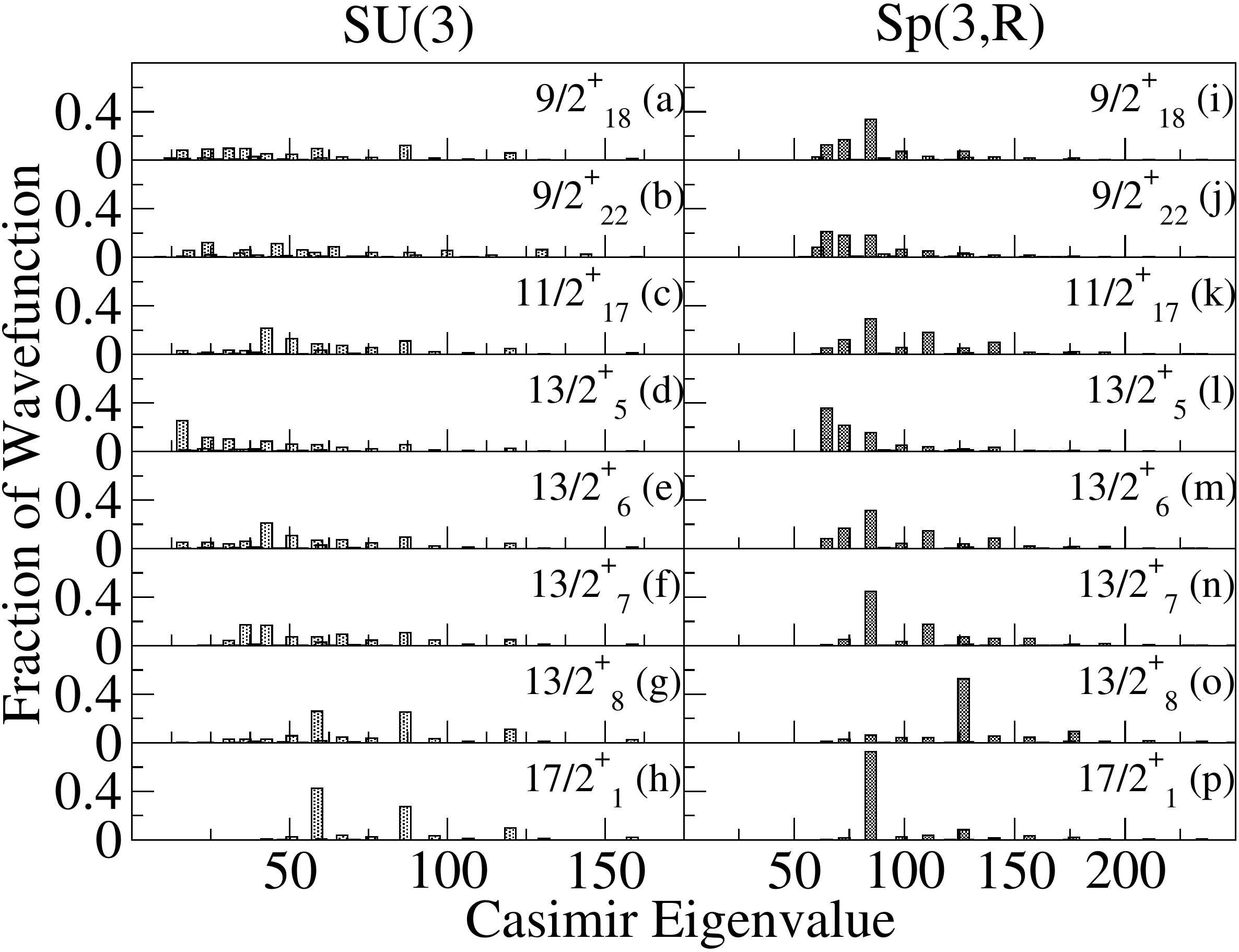}
	\caption{Decomposition of the upper rotational band members of $^9$Be found in Fig.~\ref{fig:Be9Be2}. Left-hand panels, (a)-(h): Decomposition by the quadratic Casimir of $\mathrm{SU}(3)$. Right-hand panels, (j)-(p): decomposition by the quadratic Casimir of $\mathrm{Sp}(3,R)$. Decompositions in panels (h),(p) are the same
	as Fig.~\ref{fig:Be9decomp_a}(i),(r), added here for comparison.}
	\label{fig:Be9decomp_b}
\end{figure}

Of course, $^8$Be is unbound, so that neither the $6^+$ nor $8^+$ states have been experimentally 
detected, and the continuum likely plays a role  we leave unexplored. 
Instead, we emphasize here the similarities  to $^7$Be.


\subsection{$\isotope[9]{Be}$}

An early application of $\mathrm{SU}(3)$ to light nuclei was to $^9$Be
\cite{adler1966application}. While the negative parity ground-state band is
$K=3/2$ \cite{millener2001structure}, the first excited state is the bandhead
of an unnatural positive parity $K=1/2$ band, as explained by projected
Hartree-Fock calculations \cite{bouten1968even}. Given the $2\alpha$ structure
of $^8$Be, it is reasonable to picture $^9$Be as two $\alpha$ particles plus an
extra orbiting neutron \cite{okabe1977structure}.  Although the first NCSM
calculations of $^9$Be \cite{PhysRevC.71.044312} did not focus on the
rotational motion, three rotational bands have been documented in the
calculated spectra of $^9$Be \cite{caprio2013emergence,PhysRevC.91.014310,caprio2015collective,caprio2020probing}. Two
of these bands are in the natural (negative) parity space, and the third is
within the unnatural (positive) parity space. To emphasize the generality of our findings,
 we study here the unnatural parity states shown in
Fig.~\ref{fig:Be9Be2}. 
{Once again we see what appear to be upper and lower band-like structures 
meeting at high angular momentum $J$. 
The lower $K=1/2$ band is well isolated from higher-lying
states and shows the characteristic Coriolis staggering.}

We carry out our positive parity calculations in an $N_\mathrm{max}=9$ space.
However, the natural parity $3/2^-$ ground state must be calculated in an even
$N_\mathrm{max}$ space, so, to obtain the zero point for excitation energies in
Fig.~\ref{fig:Be9Be2}, we average the ground-state energies obtained in
$N_\mathrm{max}=$ $8$ and $10$ calculations. The calculated excitation
energies for positive parity states at $N_\mathrm{max}=9$ are still significantly
higher than the experimental energies from \cite{TILLEY2004155} (also shown in
Fig.~\ref{fig:Be9Be2}), much as in other calculations \cite{PhysRevC.71.044312}.  However,
these energies are expected to continue to converge downward towards the experimental
energies with increasing $\Nmax$~\cite{caprio2019:bebands-ntse18,caprio2020probing}.
  
Although all the positive parity states are outside the natural-parity valence subspace, 
the $\Nex =1$ space plays the analogous role to the valence subspace for unnatural-parity states, and has a maximum angular momentum of $13/2$.   The {identified lower} band members 
with $J \leq 13/2$ all have about a 40-50$\%$ contribution from 
$\Nex=1$.  The band members with $J > 13/2$ (the $15/2^+_1$ and $17/2^+_1$ states), which 
contain no $\Nex=1$ configurations, 
have about $40\%$ of their wave function in the $\Nex =3$ space.

The decompositions in Fig.~\ref{fig:Be9decomp_a} presents a clear cut contrast
between $\mathrm{SU}(3)$ and $\mathrm{Sp}(3,R)$. Not surprisingly, as with our
prior beryllium cases, we observe a strong difference between $\mathrm{SU}(3)$
decompositions within and without the maximal angular momentum, in
Fig.~\ref{fig:Be9decomp_a} (h),(i).  $\mathrm{Sp}(3,R)$ reveals a nearly
perfect dynamical symmetry throughout the entire lower band, {including the 
$15/2^+_1$ and $17/2^+_1$ states.}

While we see strong transitions to an upper band in
Fig.~\ref{fig:Be9Be2}, the situation is nonetheless more complicated than we found in $^{7,8}$Be, driven at least in part, 
we believe, by mixing caused by  the high density of states.
For example, there
are strong $\Delta J =2$ $E2$ transitions between the $17/2^+_1$ state and at least four $13/2^+$ states; these states,   the $13/2^+_5$, $13/2^+_6$, $13/2^+_7$ and $13/2^+_8$ states, which in turn have strong $E2$ transitions  to the $9/2_1$,
along with  strong  transitions  from the $13/2^+_6$ state to the $9/2^+_{22}$ state, 
and from $13/2^+_7$ to $9/2^+_{18}$. 
The high density of states made it impractical to search for members of the band with $J < 9/2$.

Our decompositions, shown in Fig.~\ref{fig:Be9decomp_b}, provide additional evidence for strong mixing. (The decomposed states
correspond to filled squares in Fig.~\ref{fig:Be9decomp_b}.) 
While $^{7,8}$Be showed rather clean decompositions, $^9$Be shows much more fragmented structures. 
The SU(3) decompositions,  Fig.~\ref{fig:Be9decomp_b}(a)-(g), in particular 
are highly fragmented and have little overlap with the SU(3) decomposition of the lower band in Fig.~\ref{fig:Be9decomp_a}(a)-(i).

The Sp(3,$R$) decompositions of these excited states,
Fig.~\ref{fig:Be9decomp_b} (k)-(n), are less fragmented than their 
SU(3) counterparts, yet  are still much more fragmented than our previous Sp(3,$R$) decompositions. 
The dominant Sp(3,$R$) component found in the lower band members 
is either dominant or subdominant in all of these upper band states,
with the exception of the $13/2_{8}^+$ state. Curiously, for 
the $13/2_{8}^+$ state, the SU(3) decomposition matches those of 
the $15/2_1^+$ and $17/2_1^+$ states; all three of these states are 
predominantly $N_\mathrm{ex} =3$ and 5. To aid comparison, we duplicate the decompositions of 
the $17/2_1^+$ state from Fig.~\ref{fig:Be9decomp_a}(i),(r)  in Fig.~\ref{fig:Be9decomp_b}(h),(p).

To summarize this section, $^9$Be provides an interesting variation on our other examples. Here 
the $\Nex=1$ 
or $1 \hbar \omega$ space functions like a valence subspace for the unnatural-parity states.  The $\Nex=1$ space 
also has a maximally allowed angular momentum, and states with angular momentum $J$ greater than 
that maximally allowed value have components with $\Nex \geq 3$.  As one follows the unnatural parity 
rotational band to larger angular momentum, the $\mathrm{SU}(3)$ decomposition changes as the states exit the 
$\Nex=1$ space.  But, similar to other cases, the $\mathrm{Sp}(3,R)$ decompositions along the lower band remain 
nearly identical. While we find  strong $E2$ transitions to 
 high-lying states reminiscent of the upper bands found in $^{7,8}$Be, unlike $^{7,8}$Be we find much more 
fragmented decompositions of these states.
 Most  of these states have a significant contribution from the dominant symplectic irrep of the 
 lower band, with the exception of the $13/2^+_8$ state.

\begin{figure}[h]
\centering
	\includegraphics[width=0.9\linewidth]{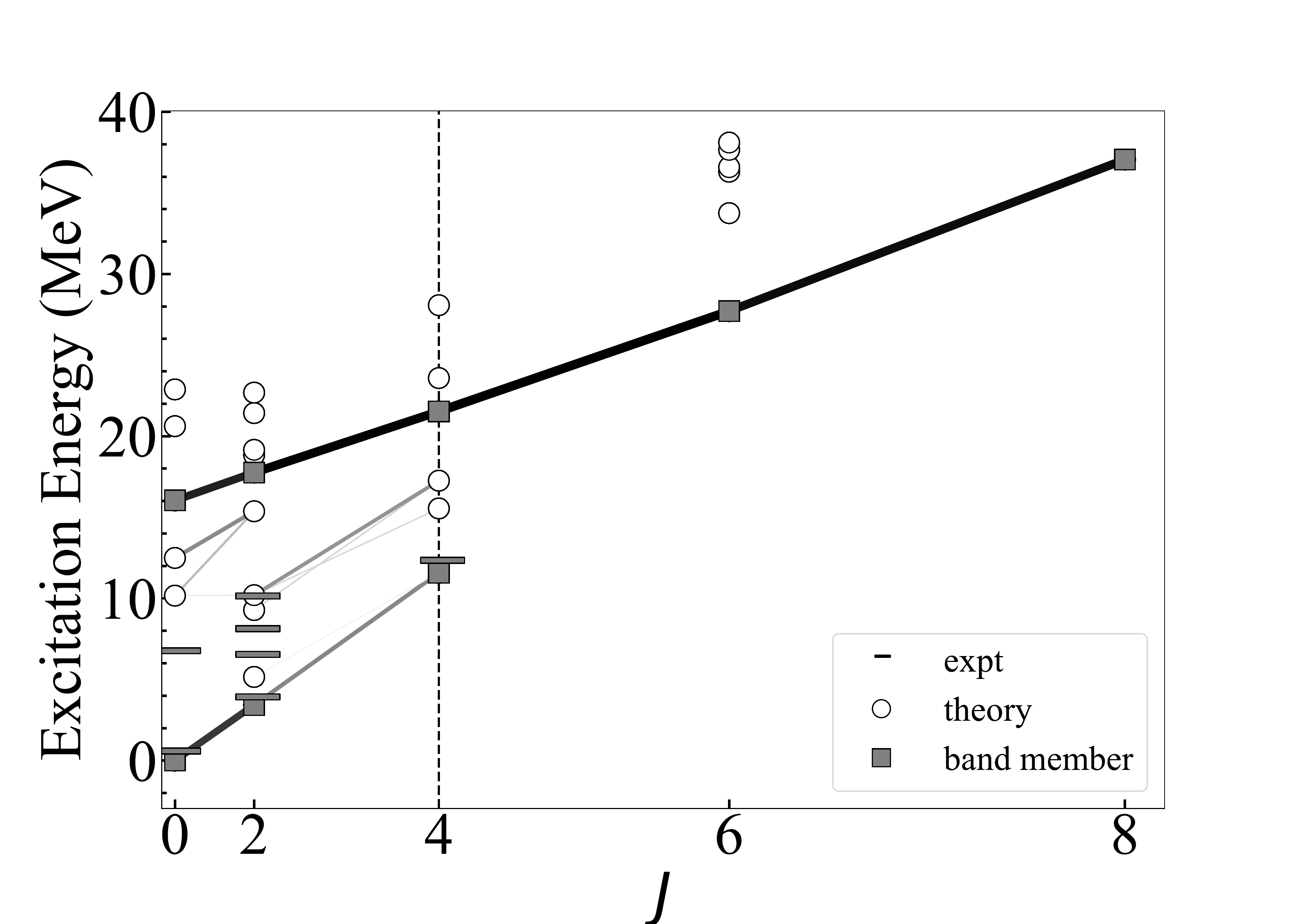}
	\caption{Excitation spectrum of the natural, positive parity states of
          ${}^{10}$Be, computed at $N_\mathrm{max} =8$ with a single-particle
          harmonic oscillator basis frequency $\hbar \omega = 20$ MeV, using a
          two-body chiral force at N3LO. Experimental energies are presented as
          shaded bars for comparison (see text). Rotational band members are
          indicated using shaded squares. These states are decomposed into group irreps in
          Fig.~\ref{fig:Be10decomp_a}.
          Line thickness
          and shading is proportional to $B(E2)$ strength. Only strengths above
          0.2 $e^2\mathrm{fm}^4$ are shown. The vertical dashed line denotes the
          end of the maximum angular momentum which can be constructed in the
          $\Nex=0$ space. }
	\label{fig:be10Be2}
\end{figure}
\begin{figure} [ht]
\centering
\begin{tabular}{c}

	\includegraphics[width=0.7\linewidth]{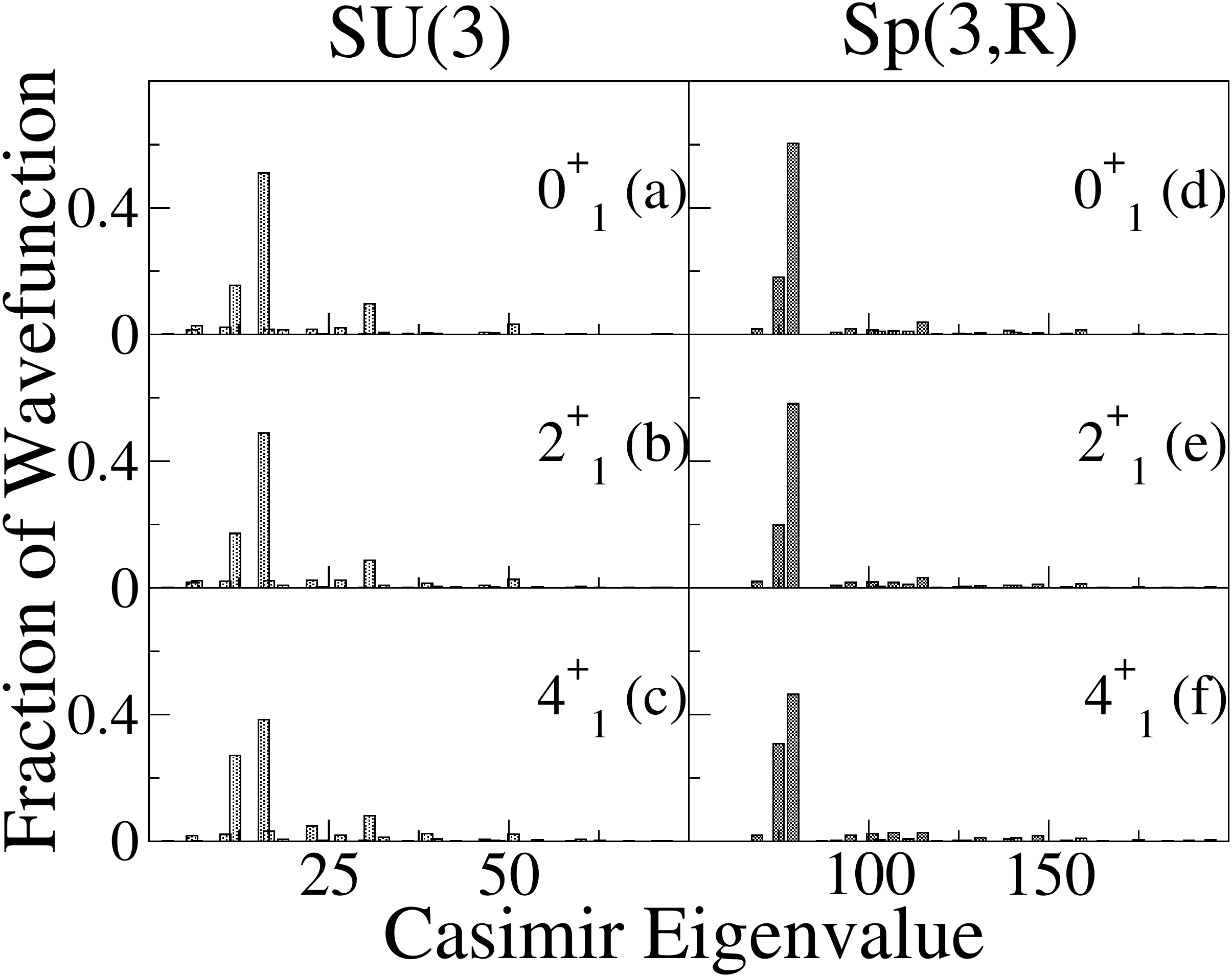}
	\\ 
		\includegraphics[width=0.7\linewidth]{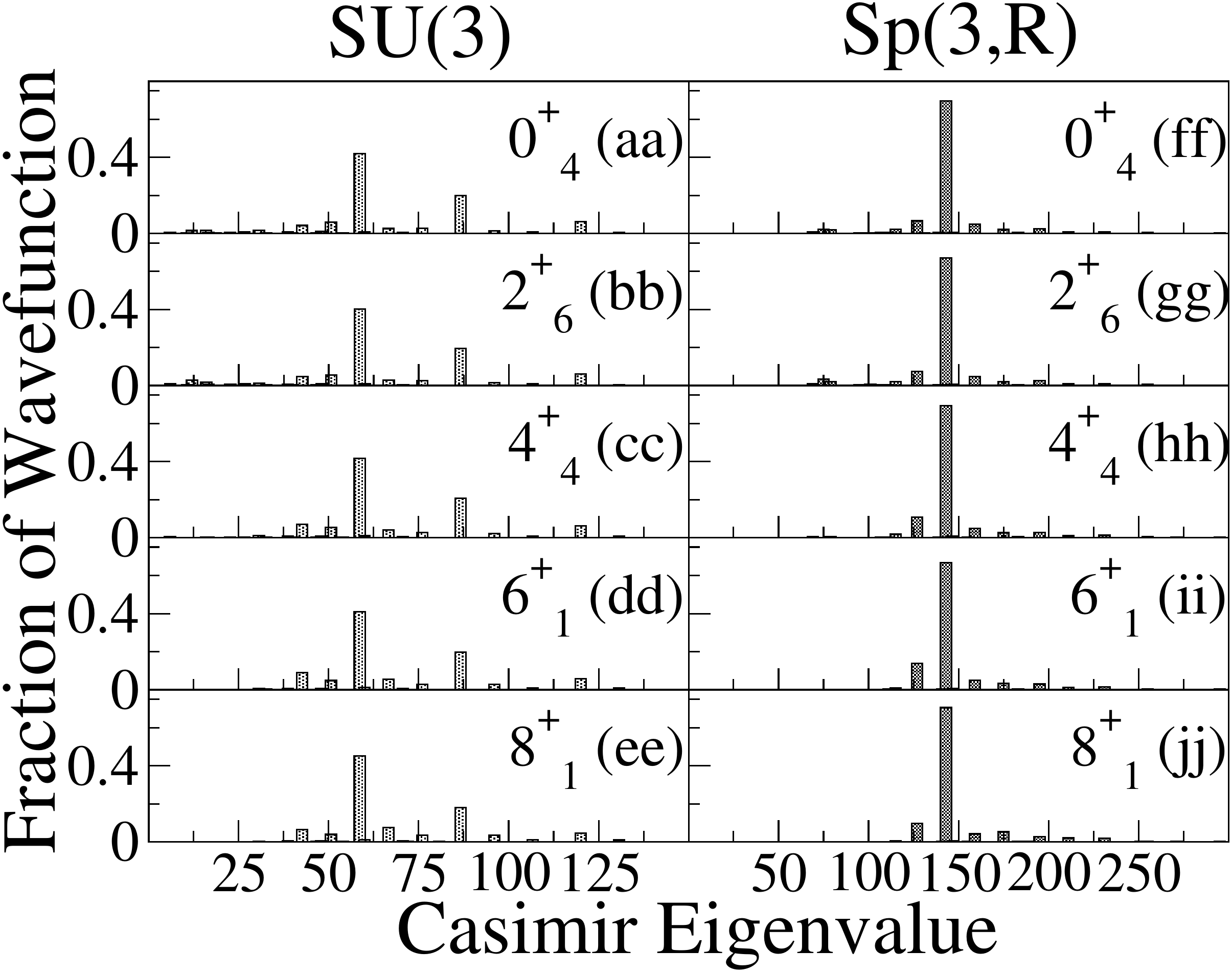}
\end{tabular}

\caption{Group-theoretical decompositions by the quadratic Casimir of $\mathrm{SU}(3)$ and $\mathrm{Sp}(3,R)$ of $^{10}$Be band members as calculated in Fig.~\ref{fig:be10Be2}. ground-state decompositions are shown in panels (a)-(f). Excited band decompositions are shown in panels (aa)-(jj).}

	\label{fig:Be10decomp_a}
\end{figure}

\medskip 

\subsection{$\isotope[10]{Be}$}

One can think of $^{10}$Be as two $\alpha$ clusters with two valence neutrons in
molecular orbitals
\cite{kanadaenyo1999:10be-amd,lashko2017microscopic}. Experimentally, $^{10}$Be
contains two distinct $K=0$ rotational bands
\cite{PhysRevLett.96.042501,PhysRevC.75.054604,PhysRevC.87.054301}, which are also
expected from, \textit{e.g.}, antisymmetrized molecular dynamics~\cite{kanadaenyo1999:10be-amd} and NCSM
calculations~\cite{caprio2013emergence,PhysRevC.91.014310,caprio2015collective,caprio2019ab}.
In the cluster picture, the molecular orbitals for the valence neutrons can be
loosely related either to spherical shell model $p$-shell orbitals for the
ground-state band, or $sd$-shell orbitals for the excited band.  In the NCSM, it
is therefore natural that these bands are built predominantly from $\Nex=0$ and
$\Nex=2$ many-body configurations, respectively~\cite{caprio2019ab}.

In Fig.~\ref{fig:be10Be2} we give 
our $N_\mathrm{max}=8$ calculation of the rotational spectra, 
along with known experimental levels \cite{TILLEY2004155}. 
In addition to an yrast band consisting of only three states
($0^+_1$, $2^+_1$, $4^+_1$),  there is also an excited band with the $0^+_4$ state as band head,
which extends beyond the maximal angular momentum $J=4$ in the
valence subspace.  Strong $E2$ transitions link the $6^+_1$, $8^+_1$ not to the 
ground-state band but to the excited band. 

This observation is supported by our analysis.
Fig.~\ref{fig:Be10decomp_a}(a)-(f) decomposes the ground-state band. As all of these 
have $\sim 70\%$
of their wave functions in the $\Nex=0$ valence subspace, it is not surprising 
that they have consistent SU(3) and $\mathrm{Sp}(3R)$ decompositions. 
The decompositions of the excited band, including the yrast $6^+_1$, $8^+_1$, in Fig.~\ref{fig:Be10decomp_a}(aa)-(jj), 
are markedly different. 
Even the members of the excited band with $J \leq 4$ have valence subspace fraction
$\leq 5\%$; instead, they consist primarily of $\Nex=2$ configurations.
Both bands exhibit clear evidence of quasi-dynamical symmetry, and given that 
the ground-state band is primarily within the valence subspace, and the excited 
band nearly exclusively outside the valence subspace, it is unsurprising that 
both SU(3) and $\mathrm{Sp}(3R)$ are good symmetries.

A similar structure involving a ground-state band terminating at low $J$ and an
excited band terminating at higher $J$ arises in NCSM calculations for
$\isotope[11]{Be}$~\cite{caprio2013emergence,PhysRevC.91.014310,caprio2020probing},
with likely counterparts in
experiment~\cite{bohlen2008:be-band,chen2019:11be-xfer}.  While we also
carried out symmetry decompositions of calculations for $\isotope[11]{Be}$, the
decompositions and resulting conclusions are comparable to those shown here for
$\isotope[10]{Be}$.  Indeed, they are sufficiently similar that a detailed presentation of
the results would be of little further illustrative value.

\begin{figure}[h]
\centering
	\includegraphics[width=0.8\linewidth]{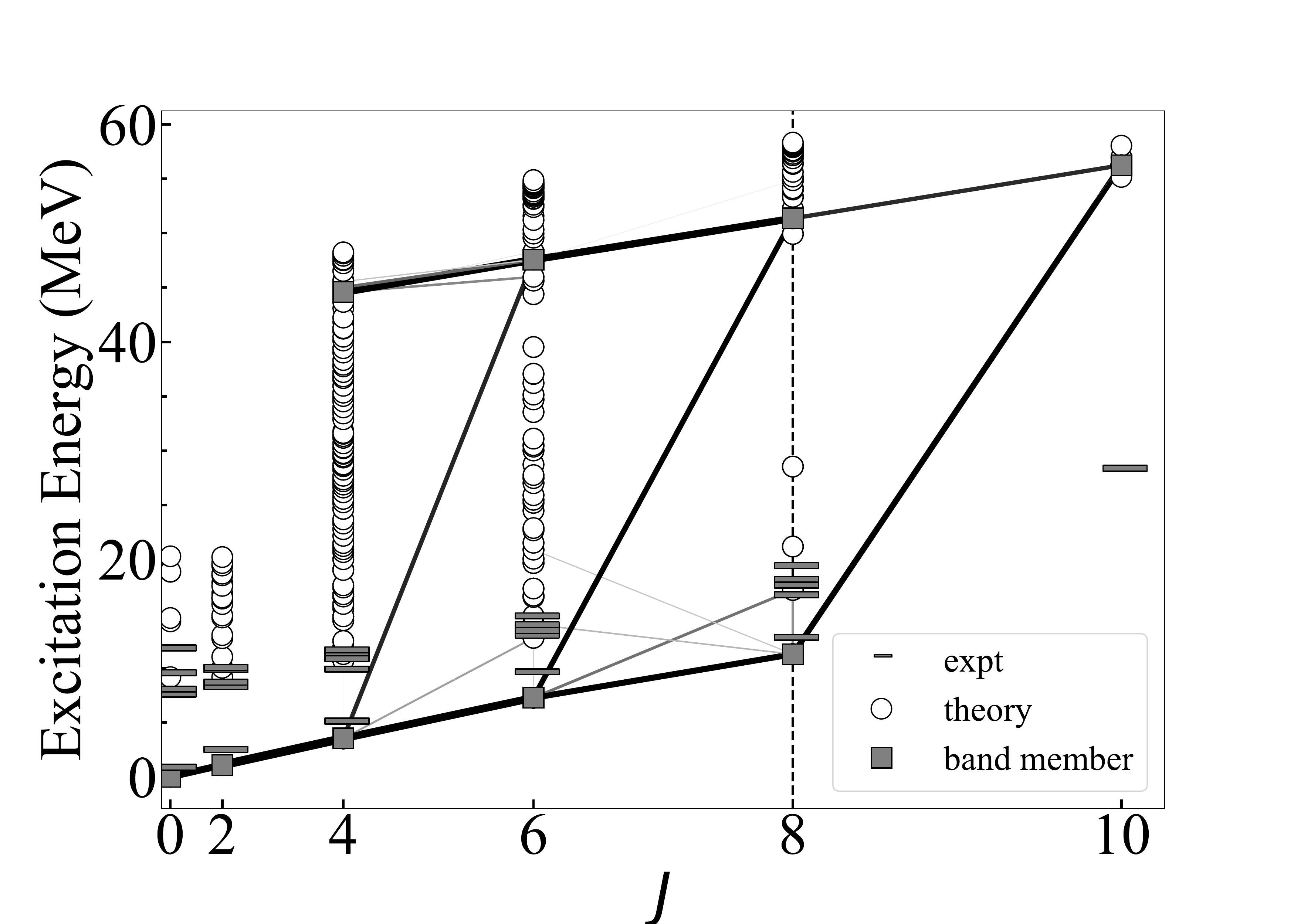}
	\caption{Excitation spectrum of the natural, positive parity
          states of ${}^{20}$Ne, restricted to even-$J$ states (see text), computed at $N_\mathrm{max} =4$ with a
          single-particle harmonic oscillator basis frequency $\hbar \omega =
          20$ MeV, using a two-body chiral force at N3LO. Experimental energies
          are presented as shaded bars for comparison (see text). Rotational
          band members are indicated using shaded squares. These states are decomposed
          into group irreps in Fig.~\ref{fig:Ne20decomp_a}. $J$-decreasing $E2$
          transitions between band members are indicated with solid connecting
          lines. Line thickness and shading is proportional to $B(E2)$
          strength. Only strengths above 0.5 $e^2\mathrm{fm}^4$ are shown.  The
          vertical dashed line denotes the maximal angular momentum within the
          lowest harmonic oscillator configuration.}
	\label{fig:Ne20Be2}
\end{figure}
\subsection{$\isotope[20]{Ne}$}

We finish with ${}^{20}$Ne, known to have an yrast rotational band
well-described by algebraic models~\cite{rosensteel1977shape}.  
This case  provides an opportunity to test the pervasiveness of our findings
by moving from the $p$ shell as valence subspace to the $sd$ shell. 
We were also motivated by a
previous study \cite{johnson2017anatomy} which found  remarkable agreement, between empirical valence subspace and \textit{ab intio} multi-shell calculations, in the $\mathrm{SU}(3)$ decomposition of both the ground-state and the first excited rotational bands.

In early calculations the 
 particle-hole 
structure of low-lying states was a subject of 
debate~\cite{abulaffio1966elliott,vogt1972rotational}; furthermore, while 
 some  particle-hole calculations failed to find strong evidence for an
extended band~\cite{goode1970ground} [that is, strong $B(E2)$ values from states
outside the valences subspace], later schematic symplectic
calculations suggested there might indeed be an extended band
\cite{ap-126-1980-343-Rosensteel,draayer1984symplectic}.

We carried out
calculations at $N_\mathrm{max}=4$,\footnote{An $\Nmax=4$, $M=0$ calculation  has a dimension of approximately 75 million, 
while an $N_\mathrm{max}=6$, $M=0$
calculation would have a dimension of 4.4 billion.  To
check the need for such a large calculation, we  calculated the $J=10$ states at
$N_\mathrm{max}=6$, using $M=10$ which has a more manageable dimension of roughly 262
million. 
We found  no significant difference between $N_\mathrm{max}=4$ and $6$ for the
SU(3) and $\mathrm{Sp}(3,R)$ decompositions of the $10^+_2$ state.} for which the excitation spectrum and
strong $E2$ transitions are shown in Fig.~\ref{fig:Ne20Be2}. 
The maximal angular momentum for ${}^{20}$Ne in the $sd$ shell is $J=8$. In
Fig.~\ref{fig:Ne20Be2} we show a ground-state band consisting of states with $J
\leq 8$.  However, the extra-valence $10^+_2$ state appears connected to the
yrast band via its strong $E2$ transition to the $8^+_1$ state.

We also observe similar strong $B(E2)$ values from the $4^+_{98}$, $6^+_{38}$,
and $8^+_6$ states to ground-state band members, as well as strong transitions among
these states indicative of band structure. (We emphasize again that the ordering
of states is specific to this model space and interaction; we do not claim, for
example, that the physical $4^+_{98}$ state is a member of the band, only that
there exists such a state.  Furthermore, we expect that our calculated
excitation energies are far from converged.)  There are no suggestions in the
calculations of any strong $E2$ transitions from these states to any candidate
odd-$J$ band members; consequently, for simplicity, only even-$J$ states are
shown in Fig.~\ref{fig:Ne20Be2}.  Thus, the excited band is identified as a
$K=0$ band, even though, due to computational limitations, we are unable extend
the excited band downward in angular momentum to $J = 0$ and $2$.

\begin{figure}[h]
\centering
    \begin{tabular}{c}
	\includegraphics[width=0.7\linewidth]{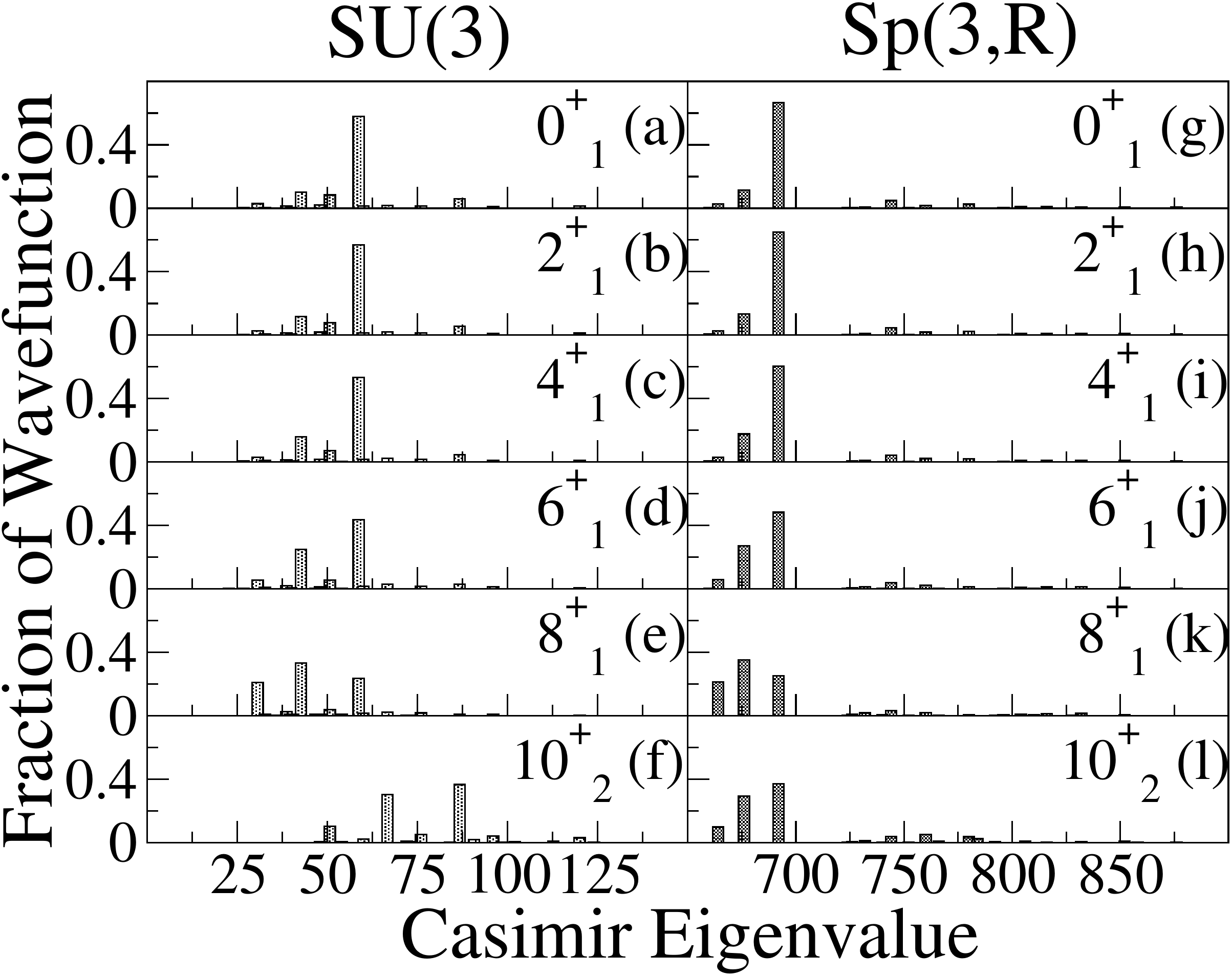}
			\\
	\includegraphics[width=0.7\linewidth]{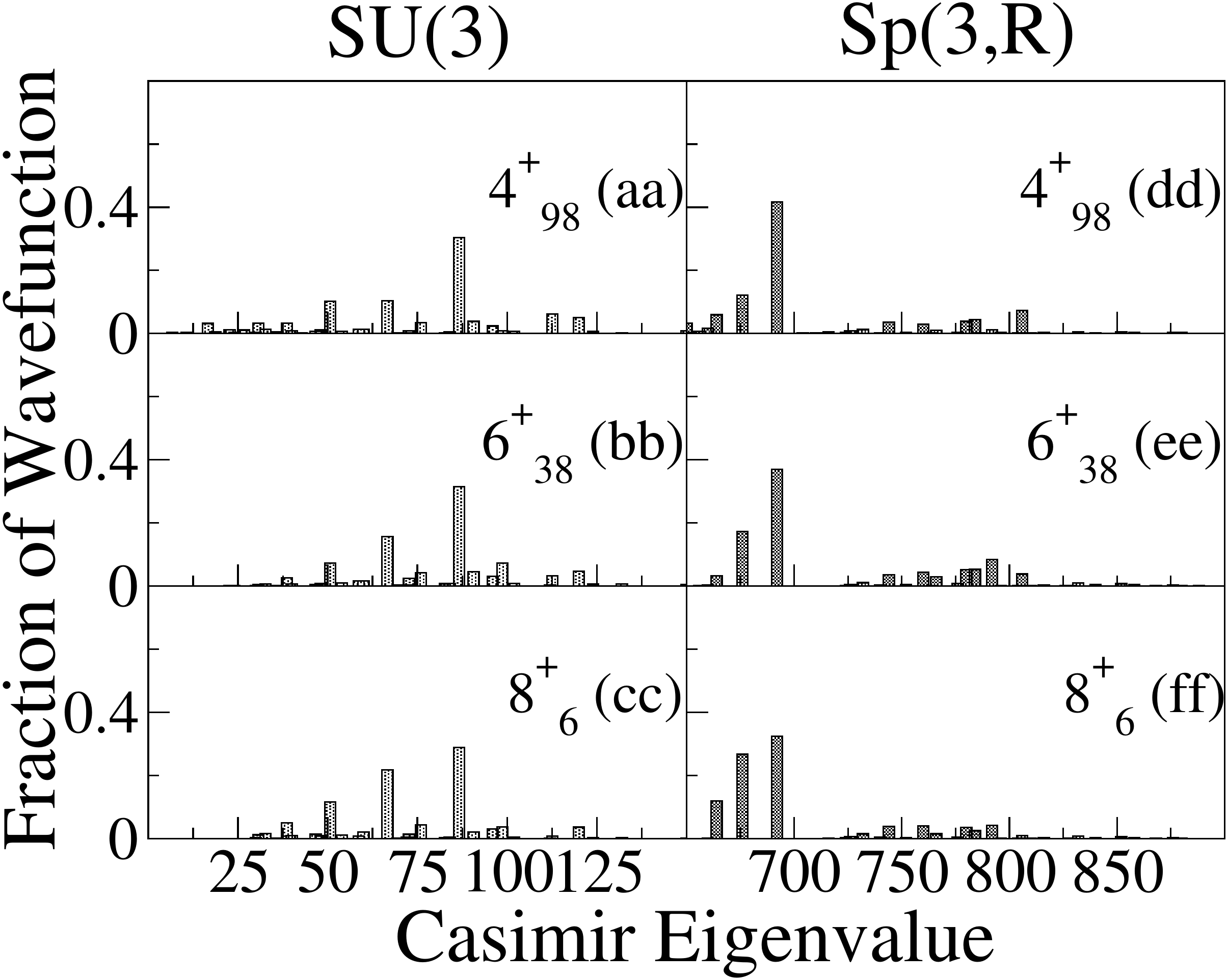}
	\end{tabular}
	
		\caption{Group-theoretical decompositions by the quadratic Casimir of $\mathrm{SU}(3)$ and $\mathrm{Sp}(3,R)$ of $^{20}$Ne band members as calculated in Fig.~\ref{fig:Be8Be2}. Ground-state decompositions are shown in panels (a)-(l). Excited band decompositions are shown in panels (aa)-(ff).}
	\label{fig:Ne20decomp_a}
\end{figure}

Members of the yrast band ($0^+_1$, $2^+_1$, $4^+_1$, $6^+_1$, $8^+_1$) have
approximately $75\%$ of their wave function in the $\Nex=0$ space, while the
states we identified as belonging to the upper band, $4^+_{98}, 6^+_{38}$,
$8^+_6$, and $10^+_2$, have approximately $70$-$80\%$ in the $\Nex=2$ space.
For yrast band members with $J \leq 8$,
the $\mathrm{SU}(3)$ decompositions in Fig.~\ref{fig:Ne20decomp_a}(a)-(e) and symplectic decompositions  
in 
Fig.~\ref{fig:Ne20decomp_a} (g)-(k) stay fairly consistent in appearance as they are all dominated 
by  $\Nex=0$ configurations. Nonetheless, we see modest evolution for decompositions with both Casimirs 
as $J$ increases, especially for the $6^+_1$, and $8^+_1$ states in  Fig.~\ref{fig:Ne20decomp_a}(d)-(e) and Fig.~\ref{fig:Ne20decomp_a}(j)-(k).  Prior work suggested  the $6^+_1$, and $8^+_1$  states have a significantly different $\Nex$ fractions compared to lower angular momentum band states~\cite{goode1970ground}. As with our beryllium calculations, Fig.~\ref{fig:Ne20decomp_a}(f) shows a radical change in the $\mathrm{SU}(3)$ decomposition as we go to $J=10$, where we must have $\Nex > 0$.

By comparing the $\mathrm{Sp}(3,R)$
decompositions for the lower band in Fig.~\ref{fig:Ne20decomp_a}(g)-(l) with
those for the upper band in Fig.~\ref{fig:Ne20decomp_a}(dd)-(ff), one can see
$^{20}$Ne has an upper band sharing the same symplectic structure.  We were not
expecting such a clear connection: Fig.~\ref{fig:Be8decomp_a}(ee) and
Fig.~\ref{fig:Be9decomp_b}(h)-(n) for the beryllium examples demonstrate the $\mathrm{Sp}(3,R)$
group decomposition can fragment for highly-excited states, which we attribute
to mixing with states nearby in energy.  Here, for $^{20}$Ne, however, despite a high density
of states the quasi-dynamical symmetry is nonetheless clear. 
Likewise, 
the $\mathrm{SU}(3)$ decompositions of $4^+_{98}$, $6^+_{38}$, and $8^+_6$ as
seen in Fig.~\ref{fig:Ne20decomp_a}(aa)-(cc) are strikingly similar to one
another and to the 
 $10^+_2$ state [Fig.~\ref{fig:Ne20decomp_a}(f)]. This is not surprising, as they are all 
 dominated by $\Nex=2$.
 
$^{20}$Ne provides strong evidence that the symplectic two-band structure found in beryllium 
nuclides extends to higher $A$. As methods and computing power continue to grow, it will be 
interesting to see how far this persists.

\section{Conclusions}

Rotational bands, identified empirically through excitation energies and strong
$E2$ transitions, are commonplace in atomic nuclei and have a long history of
being modeled algebraically and described in the context of dynamical symmetries. Of these symmetry frameworks, Elliott's
$\mathrm{SU}(3)$ is the most prominent, in part because it is designed to fit
neatly into restricted valence model spaces.  But extended bands, that is,
rotational bands that cannot be (or are not mostly) contained within the valence
subspace (for example, having an angular momentum larger than is possible within
the valence subspace), fall outside the purview of Elliott's $\mathrm{SU}(3)$
model.
Making no assumptions of the underlying group structure, we decomposed no-core
shell-model nuclear wave functions into subspaces related to the irrep labels, specifically, defined by
$\mathrm{SU}(3)$ and $\mathrm{Sp}(3,R)$ quadratic Casimirs. Such
decompositions allow one to visually compare similarities and differences in
the structure of these wave functions using  different algebraic lenses.

While valence subspace members of the 
ground-state bands follow expectations, that is,  have similar $\mathrm{SU}(3)$ decompositions, those decompositions 
differ dramatically from those of members of the extended band. 
By contrast,  members of an identified rotational 
band tend to have very similar $\mathrm{Sp}(3,R)$ decompositions.

Following  previous work \cite{mccoy2018:diss,mccoy2020emergent},
we found in addition to the usual ground state or lower band a
characteristic upper band that appears to unite with the lower band at high angular
momentum; indeed, we view these states as part of a unified feature rather than as 
physically distinct bands.  The upper and lower bands have the same symplectic decomposition,
distinct from decomposition of other nearby excited states. 
This story is not simple. We did not find an upper band with the same symplectic decomposition in 
$^{10}$Be, only a band with entirely different $\grpsu{3}$ and $\grpsptr$ structure, but it is possible we did not go up high enough in energy.
  We  also considered a system  outside 
the usual valence subspace, unnatural parity rotational bands  in $^9$Be, where 
the $\Nex=1$ or $1 \hbar \omega$ space plays a similar role to the valence subspace. While the lower 
positive parity band of $^9$Be exhibits the same behavior as our other cases,
the upper band has a much greater fragmentation of both SU(3) and Sp(3,$R$) decompositions.

Nonetheless,
we also found a unified symplectic two-band structure 
 in our calculation of $^{20}$Ne. Because of the excitation energies
involved and the fact that the generators of $\mathrm{Sp}(3,R)$ include both
quadrupole and monopole raising operators, we speculate that the upper band may
be related to giant quadrupole or monopole resonances.  This we leave as a topic
for future investigations.

\ack
This material is based upon work supported by the
U.S. Department of Energy, Office of Science, Office of Nuclear Physics, under
Award Numbers DE-FG02-00ER41132, DE-FG02-03ER41272, and DE-FG02-95ER-40934.
This research used computational resources of the National Energy Research
Scientific Computing Center (NERSC), a U.S.~Department of Energy, Office of
Science, user facility supported under Contract~DE-AC02-05CH11231.  TRIUMF
receives federal funding via a contribution agreement with the National Research
Council of Canada.

\appendix

\section{Robustness of results}

\label{robustness}

\begin{figure}[ht]
\centering
\begin{tabular}{c}
	\includegraphics[width=0.7\linewidth]{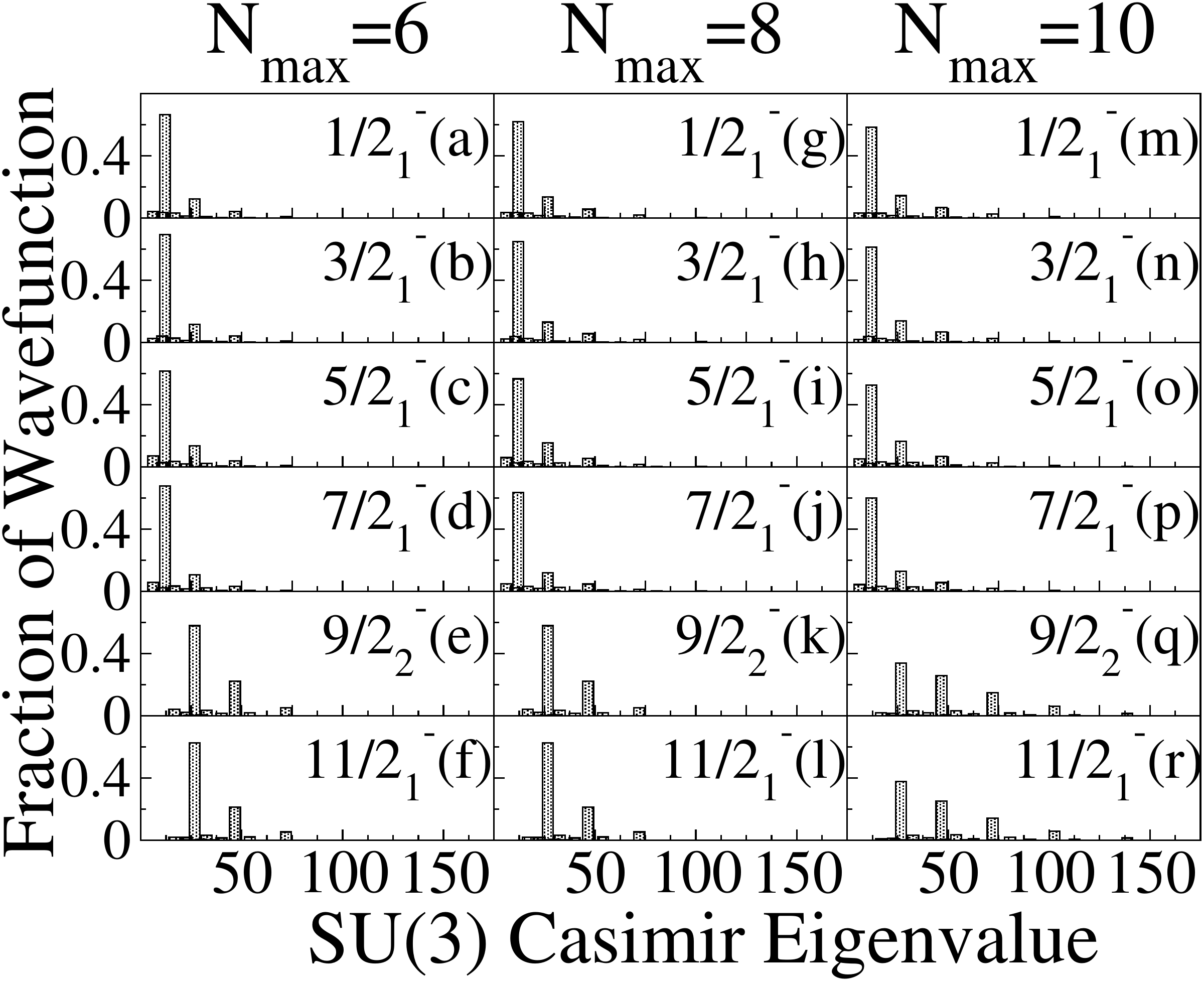} 
	\\
	\includegraphics[width=0.7\linewidth]{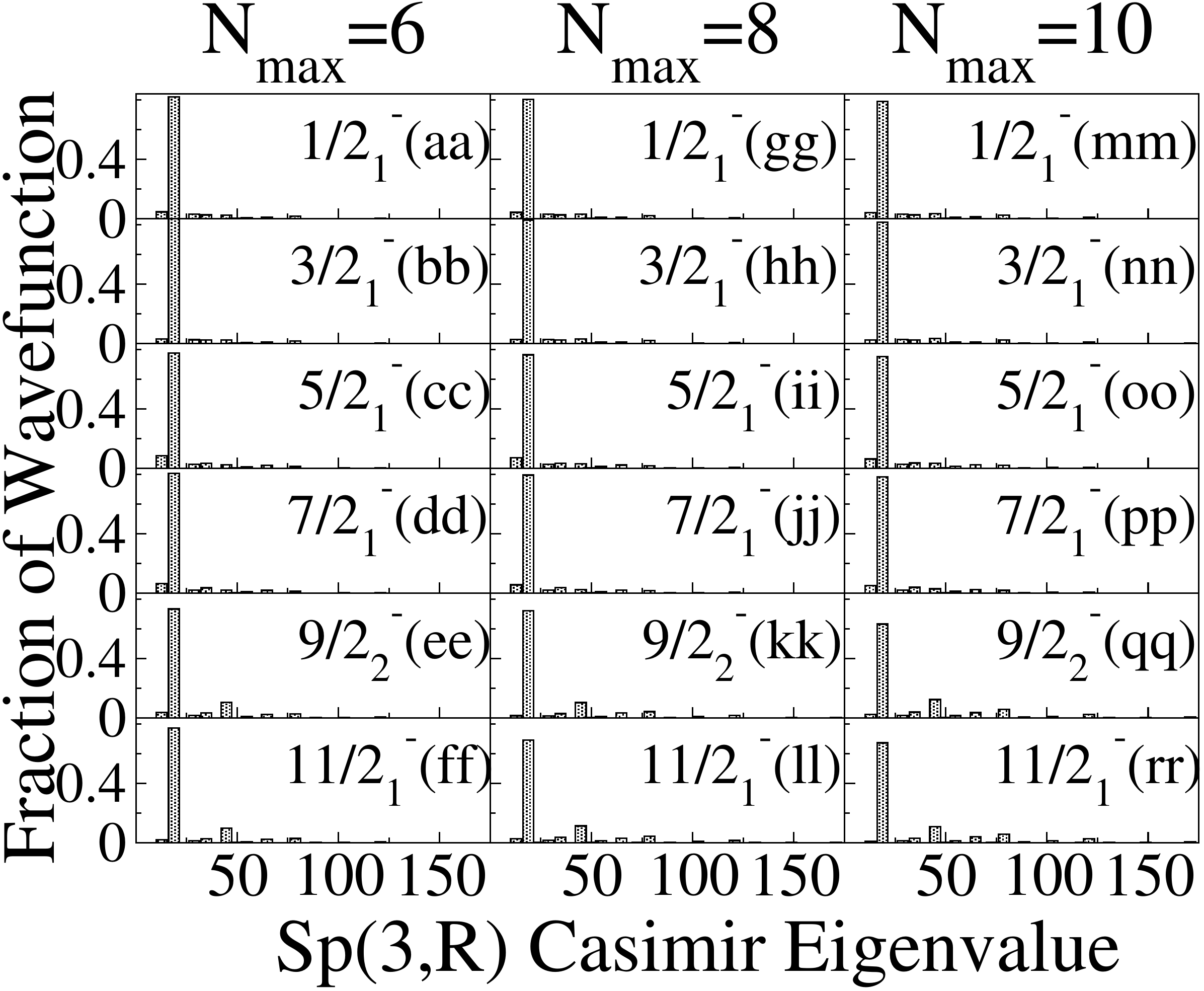} 
\end{tabular}
	\caption{Decompositions of the ground-state band of $^7$Be as calculated in 
	Fig.~\ref{fig:be7be2}. Panels (a)-(r): Decomposition by the quadratic Casimir of $\mathrm{SU}(3)$ with different $N_\mathrm{max}$ truncations. Panels (aa)-(rr): Decomposition by the quadratic Casimir of $\mathrm{Sp}(3,R)$ with different $N_\mathrm{max}$ truncations. }
	\label{fig:be7vNmax}
\end{figure}

\begin{figure}[ht]
 \centering
\begin{tabular}{c}
	\includegraphics[width=0.7\linewidth]{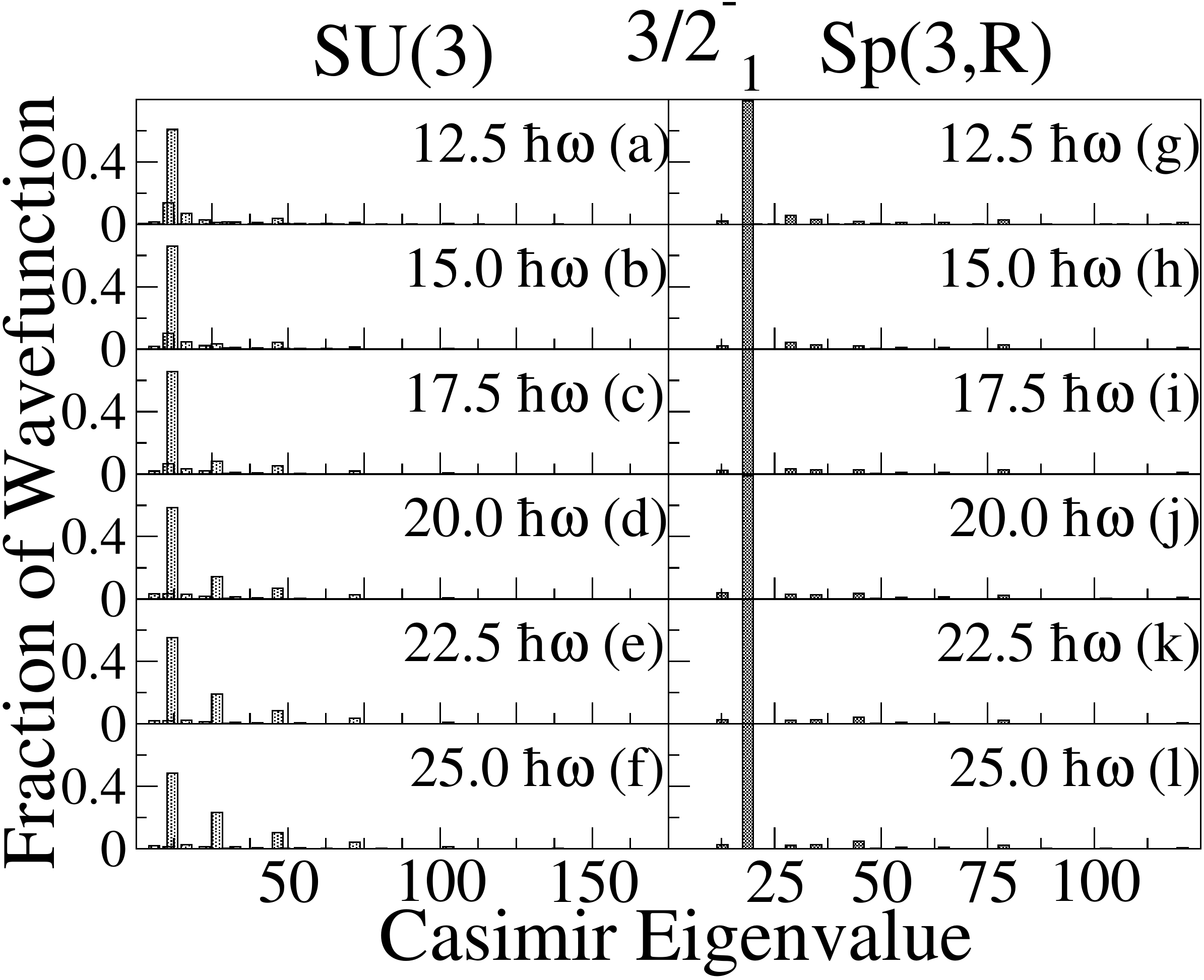} 
	\\
	\includegraphics[width=0.7\linewidth]{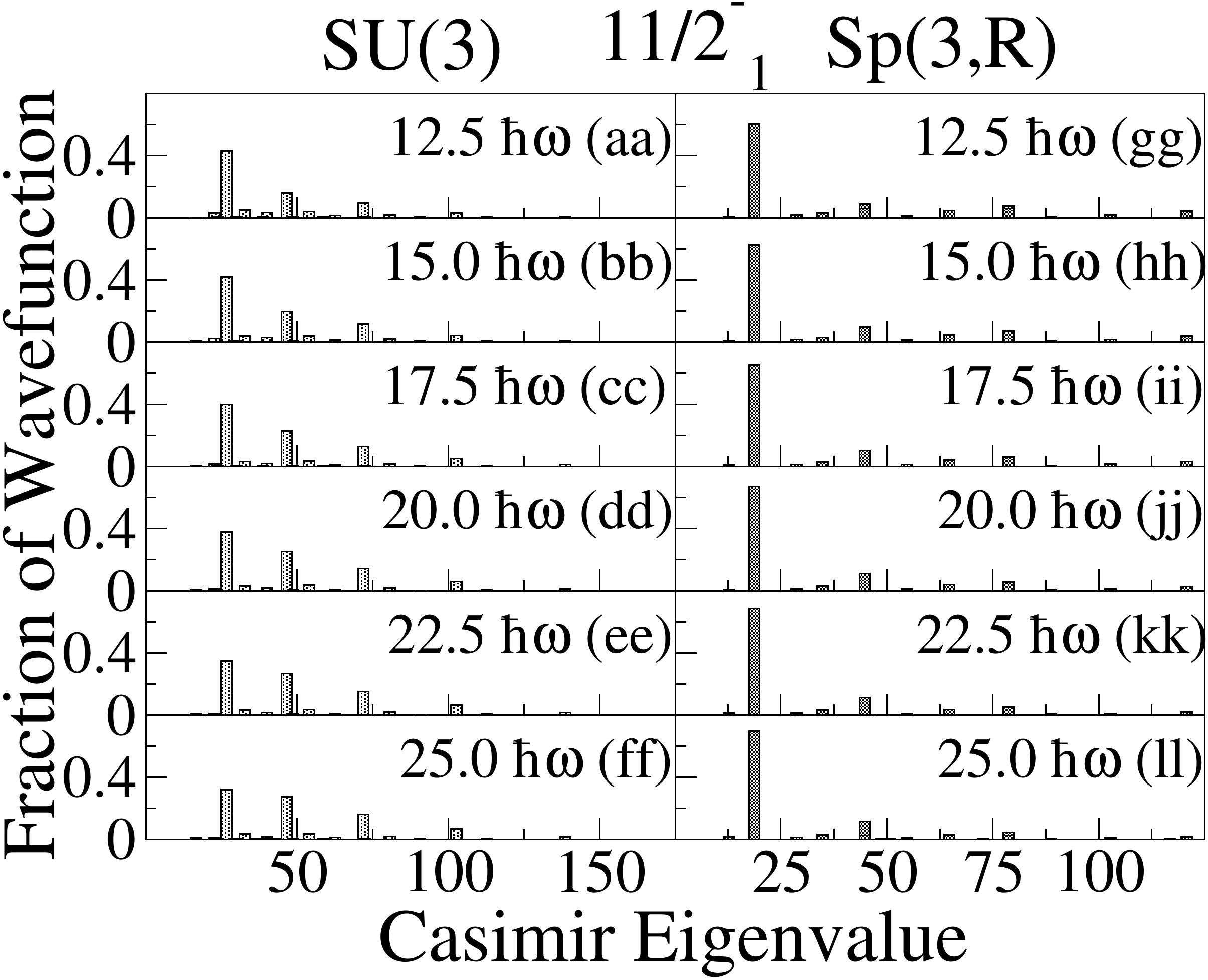} 
\end{tabular}
	\caption{Group-theoretical $\mathrm{SU}(3)$ and $\mathrm{Sp}(3,R)$ decompositions of the $J=3/2^-$ ground state, and first excited $J=11/2^-$ state of $^7$Be computed at $N_\mathrm{max}=10$ using a two-body chiral force at N3LO interaction. Each decomposition panel is labeled by the single-particle harmonic oscillator basis frequency value used to create it. Ground-state decompositions are shown in panels (a)-(l). Decompositions of the $11/2^-_1$ state are shown in panels (aa)-(ll). }

	\label{fig:be7vHw}
\end{figure}

Typically in NCSM calculations, one studies the convergence as a function of the
model space truncation $N_\mathrm{max}$ and of the basis oscillator frequence
$\hbar \omega$.  To address the question of whether or not such a study is
crucial for our purposes, we investigated the robustness of our decomposition results, specifically for $\isotope[7]{Be}$.  In
Fig.~\ref{fig:be7vNmax} we present decompositions of the ground-state band of
$^7$Be for $N_\mathrm{max}=6,8,10$, while in Fig.~\ref{fig:be7vHw} we decompose
 the $3/2^-_1$ ground state and the excited
$11/2^-_2$ state of $^7$Be, for basis $\hbar\omega$ ranging from $12.5$ to 25 MeV. The
definitions of the $\mathrm{SU}(3)$ and $\mathrm{Sp}(3,R)$ generators, and thus
Casimir operators, depend upon a choice for the harmonic oscillator length
parameter $b$ (and thus, equivalently, $\hbar\omega$), as may be seen for
$\hat{\mathcal{Q}}$ in~(\ref{ElliottQ}) (see, \textit{e.g.}, appendix~A of
reference~\cite{caprio2020:intrinsic} for detailed definitions).  The symmetry
decompositions by use of the Casimir operator may therefore depend upon this
choice as well.  In all present calculations, the generators and thus the Casimirs are defined using the
same $\hbar \omega$ as the NCSM oscillator basis.

In both cases, that is, varying both $N_\mathrm{max}$ (Fig.~\ref{fig:be7vNmax}) and $\hbar \omega$ (Fig.~\ref{fig:be7vHw}), the decompositions are remarkably robust, with slight evolution
primarily in the $\mathrm{SU}(3)$ decomposition, and most strongly in members of
the extended band.  These extra-valence states spread across $\mathrm{SU}(3)$
irreps as $N_\mathrm{max}$ increases, which accords with our understanding of
these states as being slower to converge than states that reside largely within
the valence space. In addition we see modest if understandable evolution with
$\hbar\omega$: if a state were, for example, wholly contained within the valence
space at a specific basis frequency, changing $\hbar \omega$ would mix
single-particle states with those in higher oscillator shells (in particular,
states of the same angular momentum quantum numbers $l$ and $j$, but different
nodal quantum numbers $n$), thereby changing the $\Nex$ and thus $\grpsu{3}$ decompositions. Nonetheless, the decomposition results are
remarkable consistent, that is, largely insensitive to the details of the model
space.

\section*{References}
\bibliographystyle{iopart-num}
\bibliography{master,johnsonmaster}

\providecommand{\newblock}{}
\begin{thebibliography}{10}
\expandafter\ifx\csname url\endcsname\relax
  \def\url#1{{\tt #1}}\fi
\expandafter\ifx\csname urlprefix\endcsname\relax\def\urlprefix{URL }\fi
\providecommand{\href}[2]{#1}  
\providecommand{\eprint}[2][arXiv]{#1:\linebreak[0]#2}

\bibitem{bohr1998nuclear2}
Bohr A and Mottelson B~R 1998 {\em Nuclear structure\/} vol~2 (World
  Scientific, Singapore)

\bibitem{ring2004nuclear}
Ring P and Schuck P 2004 {\em The nuclear many-body problem\/} (Springer
  Science \& Business Media, Berlin)

\bibitem{rowe2010nuclear}
Rowe D~J 2010 {\em Nuclear collective motion: models and theory\/} (World
  Scientific, Singapore)

\bibitem{rosensteel1977shape}
Rosensteel G and Rowe D 1977 {\em Annals of Physics\/} {\bf 104} 134--144

\bibitem{rowe1996dynamical}
Rowe D 1996 {\em Progress in Particle and Nuclear Physics\/} {\bf 37} 265--348

\bibitem{ui1970quantum}
Ui H 1970 {\em Progress of Theoretical Physics\/} {\bf 44} 153--171

\bibitem{rosensteel1979algebraic}
Rosensteel G and Rowe D 1979 {\em Annals of Physics\/} {\bf 123} 36--60

\bibitem{rowe2010fundamentals}
Rowe D~J and Wood J~L 2010 {\em Fundamentals of nuclear models: Foundational
  models\/} (World Scientific, Singapore)

\bibitem{talmi1993simple}
Talmi I 1993 {\em Simple models of complex nuclei\/} (CRC Press)

\bibitem{van2011scientific}
Van~Isacker P 2011 {\em Nuclear Physics A\/} {\bf 850} 157--166

\bibitem{elliott1958:su3-part1}
Elliott J~P 1958 {\em Proc. R. Soc. London\/} A {\bf 245} 128

\bibitem{elliott1958:su3-part2}
Elliott J~P 1958 {\em Proc. R. Soc. London\/} A {\bf 245} 562

\bibitem{elliott1963:su3-part3}
Elliott J~P and Harvey M 1963 {\em Proc. R. Soc. London\/} A {\bf 272} 557

\bibitem{elliott1968:su3-part4}
Elliott J~P and Wilsdon C~E 1968 {\em Proceedings of the Royal Society of
  London A: Mathematical, Physical and Engineering Sciences\/} A {\bf 302} 509

\bibitem{harvey1968nuclear}
Harvey M 1968 The nuclear $\mathrm{SU}(3)$ model {\em Advances in nuclear
  physics\/} (Springer, Berlin) pp 67--182

\bibitem{carvalho1986symplectic}
Carvalho J, Le~Blanc R, Vassanji M, Rowe D and McGrory J 1986 {\em Nuclear
  Physics A\/} {\bf 452} 240--262

\bibitem{goode1970ground}
Goode P and Wong S 1970 {\em Physics Letters B\/} {\bf 32} 89--91

\bibitem{ap-126-1980-343-Rosensteel}
Rosensteel G and Rowe D~J 1980 {\em Ann. Phys.\/} {\bf 126} 343--370

\bibitem{draayer1984symplectic}
Draayer J, Weeks K and Rosensteel G 1984 {\em Nuclear physics. A, Nuclear and
  hadronic physics\/} {\bf 413} 215--222

\bibitem{PhysRevC.91.014310}
Maris P, Caprio M~A and Vary J~P 2015 {\em Phys. Rev. C\/} {\bf 91}(1) 014310
  \urlprefix\url{http://link.aps.org/doi/10.1103/PhysRevC.91.014310}

\bibitem{PhysRevLett.38.10}
Rosensteel G and Rowe D~J 1977 {\em Phys. Rev. Lett.\/} {\bf 38}(1) 10--14
  \urlprefix\url{https://link.aps.org/doi/10.1103/PhysRevLett.38.10}

\bibitem{PhysRevLett.98.162503}
Dytrych T, Sviratcheva K~D, Bahri C, Draayer J~P and Vary J~P 2007 {\em Phys.
  Rev. Lett.\/} {\bf 98}(16) 162503
  \urlprefix\url{http://link.aps.org/doi/10.1103/PhysRevLett.98.162503}

\bibitem{PhysRevC.76.014315}
Dytrych T, Sviratcheva K~D, Bahri C, Draayer J~P and Vary J~P 2007 {\em Phys.
  Rev. C\/} {\bf 76}(1) 014315
  \urlprefix\url{http://link.aps.org/doi/10.1103/PhysRevC.76.014315}

\bibitem{0954-3899-35-12-123101}
Dytrych T, Sviratcheva K~D, Draayer J~P, Bahri C and Vary J~P 2008 {\em Journal
  of Physics G: Nuclear and Particle Physics\/} {\bf 35} 123101
  \urlprefix\url{http://stacks.iop.org/0954-3899/35/i=12/a=123101}

\bibitem{0954-3899-35-9-095101}
Dytrych T, Sviratcheva K~D, Bahri C, Draayer J~P and Vary J~P 2008 {\em Journal
  of Physics G: Nuclear and Particle Physics\/} {\bf 35} 095101
  \urlprefix\url{http://stacks.iop.org/0954-3899/35/i=9/a=095101}

\bibitem{ppnp-67-2012-516-Draayer}
Draayer J~P, Dytrych T, Launey K~D and Langr D 2012 {\em Prog. Part. and Nucl.
  Phys.\/} {\bf 67} 516

\bibitem{PhysRevLett.111.252501}
Dytrych T, Launey K~D, Draayer J~P, Maris P, Vary J~P, Saule E, Catalyurek U,
  Sosonkina M, Langr D and Caprio M~A 2013 {\em Phys. Rev. Lett.\/} {\bf
  111}(25) 252501
  \urlprefix\url{http://link.aps.org/doi/10.1103/PhysRevLett.111.252501}

\bibitem{dytrych2016efficacy}
Dytrych T, Maris P, Launey K~D, Draayer J~P, Vary J~P, Langr D, Saule E, Caprio
  M, Catalyurek U and Sosonkina M 2016 {\em Computer Physics Communications\/}
  {\bf 207} 202--210

\bibitem{launey2016symmetry}
Launey K~D, Dytrych T and Draayer J~P 2016 {\em Progress in Particle and
  Nuclear Physics\/} {\bf 89} 101--136

\bibitem{prl-124-2020-042501-Dytrych}
Dytrych T, Launey K~D, Draayer J~P, Rowe D~J, Wood J~L, Rosensteel G, Bahri C,
  Langr D and Baker R~B 2020 {\em Phys. Rev. Lett.\/} {\bf 124} 042501

\bibitem{draayer2011ab}
Draayer J, Dytrych T and Launey K 2011 Ab initio symmetry-adapted no-core shell
  model {\em Journal of Physics: Conference Series\/} vol 322 (IOP Publishing)
  p 012001

\bibitem{mccoy2018symplectic}
McCoy A~E, Caprio M~A and Dytrych T 2018 {\em Ann. Acad. Rom. Sci. Ser. Chem.
  Phys. Sci.\/} {\bf 3} 17

\bibitem{mccoy2018:diss}
McCoy A~E 2018 {\em \textit{Ab initio} multi-irrep symplectic no-core
  configuration interaction calculations\/} Ph.D. thesis University of Notre
  Dame \urlprefix\url{https://curate.nd.edu/show/pz50gt57p16}

\bibitem{mccoy2020emergent}
McCoy A~E, Caprio M~A, Dytrych T and Fasano P~J 2020 {\em Phys. Rev. Lett.\/}
  {\bf 125}(10) 102505
  \urlprefix\url{https://link.aps.org/doi/10.1103/PhysRevLett.125.102505}

\bibitem{machleidt2011chiral}
Machleidt R and Entem D~R 2011 {\em Physics Reports\/} {\bf 503} 1--75

\bibitem{PhysRevC.63.014318}
Gueorguiev V~G, Draayer J~P and Johnson C~W 2000 {\em Phys. Rev. C\/} {\bf
  63}(1) 014318
  \urlprefix\url{http://link.aps.org/doi/10.1103/PhysRevC.63.014318}

\bibitem{PhysRevC.91.034313}
Johnson C~W 2015 {\em Phys. Rev. C\/} {\bf 91}(3) 034313
  \urlprefix\url{http://link.aps.org/doi/10.1103/PhysRevC.91.034313}

\bibitem{PhysRevC.95.024303}
Herrera R~A and Johnson C~W 2017 {\em Phys. Rev. C\/} {\bf 95}(2) 024303
  \urlprefix\url{https://link.aps.org/doi/10.1103/PhysRevC.95.024303}

\bibitem{caprio2013emergence}
Caprio M~A, Maris P and Vary J~P 2013 {\em Physics Letters B\/} {\bf 719}
  179--184

\bibitem{caprio2015collective}
Caprio M~A, Maris P, Vary J~P and Smith R 2015 {\em International Journal of
  Modern Physics E\/} {\bf 24} 1541002

\bibitem{caprio2019ab}
Caprio M~A, Fasano P~J, McCoy A~E, Maris P and Vary J~P 2019 {\em Bulg. J.
  Phys.\/} {\bf 46} 445
  \urlprefix\url{https://www.bjp-bg.com/paper.php?id=1208}

\bibitem{caprio2020probing}
Caprio M~A, Fasano P~J, Maris P, McCoy A~E and Vary J~P 2020 {\em The European
  Physical Journal A\/} {\bf 56} 1--20

\bibitem{BG77}
Brussard P and Glaudemans P 1977 {\em Shell-model applications in nuclear
  spectroscopy\/} (North-Holland Publishing Company, Amsterdam)

\bibitem{br88}
Brown B~A and Wildenthal B~H 1988 {\em {Annual Review of Nuclear and Particle
  Science}\/} {\bf 38} 29--66

\bibitem{ca05}
Caurier E, Martinez-Pinedo G, Nowacki F, Poves A and Zuker A~P 2005 {\em
  {Reviews of Modern Physics}\/} {\bf 77}(2) 427--488

\bibitem{caprio2020:intrinsic}
Caprio M~A, McCoy A~E and Fasano P~J 2020 {\em J. Phys.\/} G {\bf 47} 122001

\bibitem{navratil2000large}
Navr{\'a}til P, Vary J and Barrett B 2000 {\em Physical Review C\/} {\bf 62}
  054311

\bibitem{barrett2013ab}
Barrett B~R, Navr{\'a}til P and Vary J~P 2013 {\em Progress in Particle and
  Nuclear Physics\/} {\bf 69} 131--181

\bibitem{PhysRevC.75.061001}
Bogner S~K, Furnstahl R~J and Perry R~J 2007 {\em Phys. Rev. C\/} {\bf 75}(6)
  061001 \urlprefix\url{http://link.aps.org/doi/10.1103/PhysRevC.75.061001}

\bibitem{BIGSTICK}
Johnson C~W, Ormand W~E and Krastev P~G 2013 {\em Computer Physics
  Communications\/} {\bf 184} 2761--2774

\bibitem{johnson2018bigstick}
Johnson C~W, Ormand W~E, McElvain K~S and Shan H 2018 {\em arXiv preprint
  arXiv:1801.08432\/}

\bibitem{Whitehead}
Whitehead R~R, Watt A, Cole B~J and Morrison I 1977 {\em Adv. Nuclear Phys.
  Vol. 9, pp 123-176\/}

\bibitem{hecht1973:nuclear-symmetries}
Hecht K~T 1973 {\em Annu. Rev. Nucl. Sci.\/} {\bf 23} 123

\bibitem{wybourne1974:groups}
Wybourne B~G 1974 {\em Classical Groups for Physicists\/} (New York: Wiley)

\bibitem{chen1989group}
Chen J~Q, Ping J and Wang F 1989 {\em Group representation theory for
  physicists\/} vol~7 (World Scientific)

\bibitem{iachello1994:dynsymm}
Iachello F 1994 Algebraic theory {\em Lie Algebras, Cohomology, and New
  Applications to Quantum Mechanics\/} ({\em Contemp. Math.\/} vol 160) ed
  Kamran N and Olver P (Providence, Rhode Island: American Mathematical
  Society) p 151

\bibitem{iachello2015:liealg}
Iachello F 2015 {\em Lie Algebras and Applications\/} 2nd ed ({\em Lecture
  Notes in Physics\/} vol 891) (Berlin: Springer)

\bibitem{rowe2016:micsmacs}
Rowe D~J, McCoy A~E and Caprio M~A 2016 {\em Physica Scripta\/} {\bf 91} 033003

\bibitem{rochford1988survival}
Rochford P and Rowe D 1988 {\em Physics Letters B\/} {\bf 210} 5--9

\bibitem{PhysRevC.58.1539}
Bahri C, Rowe D~J and Wijesundera W 1998 {\em Phys. Rev. C\/} {\bf 58}(3)
  1539--1550 \urlprefix\url{http://link.aps.org/doi/10.1103/PhysRevC.58.1539}

\bibitem{rowe2000quasi}
Rowe D~J 2000 Quasi-dynamical symmetry:a new use of symmetry in nuclear physics
  {\em The Nucleus\/} (Springer) pp 379--395

\bibitem{bahri20003}
Bahri C and Rowe D 2000 {\em Nuclear Physics A\/} {\bf 662} 125--147

\bibitem{whitehead1980:lanczos}
Whitehead R~R 1980 {\em Theory and Applications of Moment Methods in
  Many-Fermion Systems\/} ed Dalton B~J, Grimes S~M, Vary J~P and Williams S~A
  (New York: Plenum) p 235

\bibitem{PhysRevC.65.054309}
Escher J and Leviatan A 2002 {\em Phys. Rev. C\/} {\bf 65}(5) 054309
  \urlprefix\url{https://link.aps.org/doi/10.1103/PhysRevC.65.054309}

\bibitem{JISP16}
Shirokov A, Vary J, Mazur A and Weber T 2007 {\em Physics Letters B\/} {\bf
  644} 33--37

\bibitem{ekstroem2013:nnlo-opt}
Ekstr{\"o}m A, Baardsen G, Forss{\'e}n C, Hagen G, Hjorth-Jensen M, Jansen G~R,
  Machleidt R, Nazarewicz W, Papenbrock T, Sarich J and Wild S~M 2013 {\em
  Phys. Rev. Lett.\/} {\bf 110} 192502

\bibitem{Daejeon16}
Shirokov A, Shin I, Kim Y, Sosonkina M, Maris P and Vary J 2016 {\em Physics
  Letters B\/} {\bf 761} 87--91

\bibitem{TILLEY20023}
Tilley D, Cheves C, Godwin J, Hale G, Hofmann H, Kelley J, Sheu C and Weller H
  2002 {\em Nuclear Physics A\/} {\bf 708} 3 -- 163 ISSN 0375-9474
  \urlprefix\url{http://www.sciencedirect.com/science/article/pii/S0375947402005973}

\bibitem{TILLEY2004155}
Tilley D, Kelley J, Godwin J, Millener D, Purcell J, Sheu C and Weller H 2004
  {\em Nuclear Physics A\/} {\bf 745} 155 -- 362 ISSN 0375-9474
  \urlprefix\url{http://www.sciencedirect.com/science/article/pii/S0375947404010267}

\bibitem{yoshida2013cluster}
Yoshida T, Shimizu N, Abe T and Otsuka T 2013 Cluster structure in {M}onte
  {C}arlo shell model {\em Journal of Physics. Conference Series (Online)\/}
  vol 454

\bibitem{bouten1970ground}
Bouten M, Bouten M, Depuydt H and Schotsmans L 1970 {\em Physics Letters B\/}
  {\bf 33} 457--459

\bibitem{bouten1971ground}
Bouten M, Bouten M and Van~Leuven P 1971 {\em Nuclear Physics A\/} {\bf 168}
  438--448

\bibitem{arickx1975configuration}
Arickx F, Van~Leuven P and Bouten M 1975 {\em Nuclear Physics A\/} {\bf 252}
  416--422

\bibitem{arickx1976new}
Arickx F 1976 {\em Nuclear Physics A\/} {\bf 268} 347--357

\bibitem{arickx1979sp}
Arickx F, Broeckhove J and Deumens E 1979 {\em Nuclear Physics A\/} {\bf 318}
  269--286

\bibitem{adler1966application}
Adler C, Corcoran T and Mast C 1966 {\em Nuclear Physics\/} {\bf 88} 145--168

\bibitem{millener2001structure}
Millener D 2001 {\em Nuclear Physics. A\/} {\bf 693} 394--410

\bibitem{bouten1968even}
Bouten M, Bouten M, Depuydt H and Schotsmans L 1968 {\em Physics Letters B\/}
  {\bf 27} 61--64

\bibitem{okabe1977structure}
Okabe S, Abe Y and Tanaka H 1977 {\em Progress of Theoretical Physics\/} {\bf
  57} 866--881

\bibitem{PhysRevC.71.044312}
Forss\'en C, Navr\'atil P, Ormand W~E and Caurier E 2005 {\em Phys. Rev. C\/}
  {\bf 71}(4) 044312
  \urlprefix\url{https://link.aps.org/doi/10.1103/PhysRevC.71.044312}

\bibitem{caprio2019:bebands-ntse18}
Caprio M~A, Fasano P~J, Vary J~P, Maris P and Hartley J 2019 Robust \textit{ab
  initio} predictions for nuclear rotational structure in the $\isotope{Be}$
  isotopes {\em Proceedings of the International Conference Nuclear Theory in
  the Supercomputing Era 2018\/} ed Shirokov A~M and Mazur A~I (Pacific
  National University, Khabarovsk, Russia) p 250
  \urlprefix\url{http://www.ntse.khb.ru/files/uploads/2018/proceedings/Caprio.pdf}

\bibitem{kanadaenyo1999:10be-amd}
Kanada-En'yo Y, Horiuchi H and Dot{\'e} A 1999 {\em Phys. Rev.\/} C {\bf 60}
  064304

\bibitem{lashko2017microscopic}
Lashko Y~A, Filippov G and Vasilevsky V 2017 {\em Nuclear Physics A\/} {\bf
  958} 78--100

\bibitem{PhysRevLett.96.042501}
Freer M, Casarejos E, Achouri L, Angulo C, Ashwood N~I, Curtis N, Demaret P,
  Harlin C, Laurent B, Milin M, Orr N~A, Price D, Raabe R,
  Soi\ifmmode~\acute{c}\else \'{c}\fi{} N and Ziman V~A 2006 {\em Phys. Rev.
  Lett.\/} {\bf 96}(4) 042501
  \urlprefix\url{https://link.aps.org/doi/10.1103/PhysRevLett.96.042501}

\bibitem{PhysRevC.75.054604}
Bohlen H~G, Dorsch T, Kokalova T, Oertzen W~v, Schulz C and Wheldon C 2007 {\em
  Phys. Rev. C\/} {\bf 75}(5) 054604
  \urlprefix\url{https://link.aps.org/doi/10.1103/PhysRevC.75.054604}

\bibitem{PhysRevC.87.054301}
Suzuki D, Shore A, Mittig W, Kolata J~J, Bazin D, Ford M, Ahn T, Becchetti F~D,
  Beceiro~Novo S, Ben~Ali D, Bucher B, Browne J, Fang X, Febbraro M, Fritsch A,
  Galyaev E, Howard A~M, Keeley N, Lynch W~G, Ojaruega M, Roberts A~L and Tang
  X~D 2013 {\em Phys. Rev. C\/} {\bf 87}(5) 054301
  \urlprefix\url{https://link.aps.org/doi/10.1103/PhysRevC.87.054301}

\bibitem{bohlen2008:be-band}
Bohlen H~G, von Oertzen W, Kalpakchieva R, Massey T~N, Dorsch T, Milin M,
  Schulz {\mbox{Ch}}, Kokalova {\mbox{Tz}} and Wheldon C 2008 {\em J. Phys.
  Conf. Ser.\/} {\bf 111} 012021

\bibitem{chen2019:11be-xfer}
Chen J, Auranen K, Avila M~L, Back B~B, Caprio M~A, Hoffman C~R, Gorelov D, Kay
  B~P, Kuvin S~A, Liu Q, Lou J~L, Macchiavelli A~O, McNeel D~G, Tang T~L,
  Santiago-Gonzalez D, Talwar R, Wu J, Wilson G, Wiringa R~B, Ye Y~L, Yuan C~X
  and Zang H~L 2019 {\em Phys. Rev.\/} C {\bf 100} 064314

\bibitem{johnson2017anatomy}
Johnson C~W 2017 The anatomy of atomic nuclei: illuminating many-body wave
  functions through group-theoretical decomposition {\em Emergent Phenomena In
  Atomic Nuclei From Large-scale Modeling: A Symmetry-guided Perspective\/} ed
  Launey K~D (World Scientific, Singapore) p~33

\bibitem{abulaffio1966elliott}
Abulaffio C 1966 {\em Nuclear Physics\/} {\bf 81} 71--75

\bibitem{vogt1972rotational}
Vogt E 1972 {\em Physics Letters B\/} {\bf 40} 345--348

\end{thebibliography}

\end{document}